\documentclass[twocolumn,showpacs,preprintnumbers]{revtex4}
\usepackage{amssymb}
\usepackage{amsfonts}
\usepackage{amsmath}
\usepackage{graphicx}
\usepackage{dcolumn}
\usepackage{bm}
\usepackage{hyperref}

\input{tcilatex}
\begin{document}

\title{Local positioning system as a classic alternative to atomic
navigation.}
\author{B. Dubetsky}
\affiliation{Independent Researcher, 1849 S Ocean Dr, Apt 207, Hallandale Florida 33009,
United States}
\email{bdubetsky@gmail.com}
\date{\today }

\begin{abstract}
A local positional system (LPS) is proposed, in which particles are launched
at given velocities, and a sensor system measures the trajectory of
particles in the platform frame. These measurements allow us to restore the
position and orientation of the platform in the frame of the rotating Earth,
without solving navigation equations. When the platform velocity is known
and if the platform orientation stays the same, the LPS-technique allows a
navigational accuracy of 100$\mu $ per one hour to be achieved . In this
case, the LPS technique is insensitive to the type of platform trajectory.
If there are also velocimeters installed on the platform, then one can
restore the velocity and angular rate of the platform rotation in respect to
the Earth. Instead of navigational equations, it is necessary to obtain the
classical trajectory of a particle in the field of a rotating gravity
source. Taking into account the gravity-gradient, Coriolis, and centrifugal
forces, the exact expression for this trajectory is derived, which can be
widely used in atomic interferometry. A new iterative method for restoring
the orientation of the platform without using gyroscopes is developed. The
simulation allowed us to determine the conditions under which the
LPS-navigation error per hour is about $10$m.
\end{abstract}

\maketitle

\twocolumngrid%

\section{\label{s1}Introduction}

Since its birth about 40 years ago \cite{c1}, the field of atom
interferometry has matured significantly. The current state and prospects in
this area are presented, for example, in the reviews \cite{c1.1} and the
proposals \cite{c1.2,c1.3,c1.4,c1.5,c1.5.1,c1.5.2}.

Among other applications, atom interferometers (AIs) are now proposed to use
as inertial measurement units (IMUs) in navigation.%

The precision accuracy of AI allows us to hope for significant progress,
improving the accuracy and duration of the navigation. These applications
are based on the possibility of using AI as an accelerometer \cite{c2}. If
the atoms move with acceleration $a$, then the AI phase is given by%
\begin{equation}
\phi =\mathbf{k}\cdot \mathbf{a}T^{2},  \label{1}
\end{equation}%
where $\mathbf{k}$ is the effective wave vector of the Raman pulses, $T$ is
the delay time between these pulses. The error of the acceleration
measurement is 
\begin{equation}
\delta a=\phi _{err}/kT^{2},  \label{2}
\end{equation}%
where $\phi _{err}$ is the accuracy of the AI phase measurement. To date,
for example, the accuracy of differential acceleration measurements has
reached 
\begin{equation}
\delta a=1.4ppt  \label{2.1}
\end{equation}%
at the delay time 
\begin{equation}
T=955ms.  \label{2.2}
\end{equation}

For navigation applications, a system of 6 AIs was created and tested \cite%
{c4}. However, the direct use of precise AI as an IMU is difficult. Matter
of fact is that the IMUs data should be substituted into the navigation
equations, and for an accurate numerical solution of these equations, the
IMUs measurements should be repeated with a time step of at least $\tau \sim
1$ms \cite{c5}. To do this, the interrogation time of the AI must be no more
than $\tau $, and therefore $T\lesssim \tau /2$. Thus, with the same
accuracy of the phase measurement $\phi _{err}$ and the effective wave
vector $k$, the acceleration measurement error increases by 6 orders of
magnitude to the level of $5$ppm. With such an inaccuracy, atomic
interferometers will be even an order of magnitude worse than conventional
accelerometers \cite{c6} and will not result in any progress in navigation.

Two methods have been proposed to avoid this difficulty. In one method, AIs
is used in a hybrid mode together with conventional IMUs \cite%
{c7,c8,c9,c9.1,c10}. Another option \cite{c11} is related to the fact that
one can restore the position of the platform directly from the AI phase.
without measuring the atomic acceleration and, in principle, without any use
of conventional IMUs. To understand why this is possible, one notes that Eq.
(\ref{1}) is valid only for the uniform accelerated motion of the atom. If
the atomic cloud moves in the platform frame, then, ignoring the quantum
effects, for the phase instead of (\ref{1}) one gets 
\begin{subequations}
\label{3}
\begin{eqnarray}
\phi &=&\mathbf{k}\cdot \mathbf{p},  \label{3a} \\
\mathbf{p} &=&\mathbf{x}\left( t_{3}\right) -2\mathbf{x}\left( t_{2}\right) +%
\mathbf{x}\left( t_{1}\right) ,  \label{3b}
\end{eqnarray}%
where $\mathbf{x}(t)$ is the classical trajectory of the cloud in the
platform frame. Knowing the trajectory at the previous moments $t_{1}$ and $%
t_{2}=t_{1}+T$ and measuring the phase of AI, one restores the trajectory at
the moment $t_{3}=t_{1}+2T$. At the same time, there is no need to solve
navigation equations, and as a result, the restriction on the feasible delay
time $T$ is removed. One sees that when using AI as an IMU, one actually
does double work: first, by measuring the trajectory, $\mathbf{x}\left(
t_{3}\right) $, gets the acceleration, and then from the navigation
equations gets the trajectory.

The error of measuring the position $\mathbf{x}\left( t_{3}\right) $ is
given by 
\end{subequations}
\begin{equation}
\delta _{x}\sim \phi _{err}/k\sim 1.3\text{pm,}  \label{4}
\end{equation}%
where to estimate the parameter $\phi _{err}/k$, one can use the Eqs. (\ref%
{2.1}-\ref{2.2}). Despite the fact that at each step the error (\ref{4})
does not depend on $T$, the total error $\Delta X$ during navigation time $%
t_{n}$ accumulates after repeated measurement of phases at times multiple to 
$T$, . For the given $t_{n}$, the number of steps

\begin{equation}
n\sim t_{n}/T.  \label{5}
\end{equation}%
The use of AIs with a longer interrogation time leads to a reduction in the
number of steps and, consequently, to a reduction in the accumulated error.
Thus, one sees that, proposed in \cite{c11}, the rejection of the use of AIs
as accelerometers, allows the use of an unprecedently high accuracy of
trajectory measurement (\ref{4}) for an unprecedently accurate navigation.

Atom interference is an exclusively quantum phenomenon. Nevertheless, the
expression (\ref{3}) for the phase is purely classical. This means that
instead of quantum objects, atoms, one can use classical material points,
which in the future we will call particles. We do not specify the kind of
particles. However, we assume that the mass of the particle $M$ is
sufficiently large to neglect the recoil velocity $\hbar k/M$, where, in our
case, $\hbar k$ is the change in momentum that accompanies the measurement
of the particle position.

Like in the patent \cite{c11}, in this article, we propose to abandon the
use of accelerometers, gyroscopes, and navigation equations, and instead
measure the trajectories of particles that are launched in the platform
system from certain points, at certain speeds, and with a time interval $T$.
We will develop a theory of such navigation here and perform numerical
simulations. The obvious advantages of classical objects are a dramatic
mitigation in the requirements of high vacuum, disappearing the necessity to
solve the problem of the phase multiple ambiguity, and the absence of a
recoil effect. The obvious disadvantage of our method is that for classical
objects, the accuracy of the measurement (\ref{4}) of their position is
hardly achievable at the present time. We have, however, developed methods
here that will significantly relax the requirements for $\delta _{x}$ The
ultimate goal of numerical simulation is to answer the question of what the
accuracy of measuring the coordinates and velocities of particles should be
in order to achieve a given accuracy of navigation $\Delta X$ for a given
period of time $t_{n}$.Specifically, we will find the answer to this
question for 
\begin{subequations}
\label{6}
\begin{eqnarray}
t_{n} &=&1\text{h,}  \label{6a} \\
\Delta X &=&10\text{m.}  \label{6b}
\end{eqnarray}

The accelerometer in conventional navigation measures the acceleration of
particle $\mathbf{a}$ in the platform system. For $\mathbf{a}$ in the
absence of rotation one has 
\end{subequations}
\begin{equation}
\mathbf{a}=\mathbf{g}-\mathbf{A},  \label{7}
\end{equation}%
where $\mathbf{g}$ and $\mathbf{A}$ are the gravitational field of the Earth
and the acceleration of the platform. Knowing $\mathbf{g}$ and measuring $%
\mathbf{a}$, one obtains the acceleration of platform $\mathbf{A}$,
substituting it into the navigation equation%
\begin{equation}
\mathbf{\ddot{X}}\left( t\right) =\mathbf{A},  \label{8}
\end{equation}%
to determine the trajectory of the platform $\mathbf{X}\left( t\right) $.

\begin{figure}[!t]
\includegraphics[width=8cm]{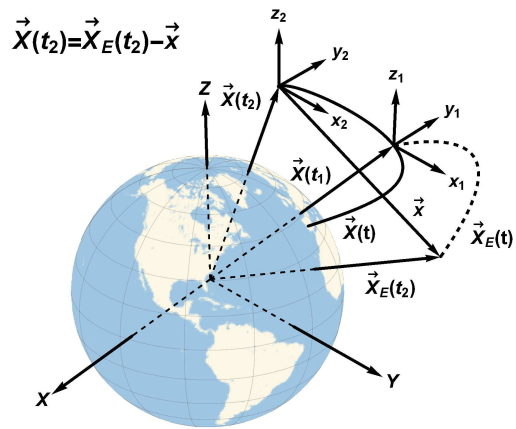}
\caption{Exploiting the measurement of the particle
position in the platform frame for. navigation. The trajectories of the
platform and the particle are shown with solid and dotted curves.}
\label{f1}
\end{figure}

In Ref. \cite{c11}, it is actually proposed to replace the accelerations $%
\mathbf{a}$ and $\mathbf{A}$, with the trajectories $\mathbf{x}(t)$ and $%
\mathbf{X}\left( t\right) $. The procedure for determining the position of
the platform by measuring the position of the particle in the absence of
rotation of the platform is illustrated in Fig. \ref{f1}.

\begin{figure}[!t]
\includegraphics[width=8cm]{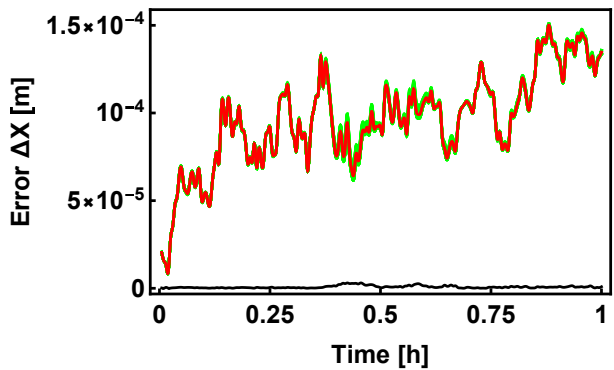}
\caption{The green curves are the time
dependences of the navigation error $\Delta X\left( t\right) $ for an
ensemble of 30 randomly generated platform trajectories $\mathbf{X}\left(
t\right) $. The red curve is the average error in the ensemble $\left\langle
\Delta X\left( t\right) \right\rangle $, the black curve is SD $\protect%
\sqrt{\left\langle \left[ \Delta X\left( t\right) \right] ^{2}\right\rangle
-\left\langle \Delta X\left( t\right) \right\rangle ^{2}}.$}
\label{f2}
\end{figure}

We are interested in the trajectory of
the platform $\mathbf{X}\left( t\right) $ in the coordinate system $\left\{
X,Y,Z\right\} $ rotating with the Earth. On the platform, one installs a
local frame $\left\{ x,y,z\right\} $ with the origin at the point $\mathbf{X}%
\left( t\right) $ and sensors that measure the position of the particle $%
\mathbf{x}(t)$ in that local frame. At the moment of time $t_{1}$ one
launches the particle. After the launch, the particle is completely
decoupled from the platform and moves along the trajectory $\mathbf{X}%
_{E}\left( t\right) $ in a vacuum under the influence of only the rotating
Earth. Knowing the potential of the Earth's gravitational field and the
velocity of the Earth rotation $\vec{\Omega}_{E}\left( t\right) ,$ one can
obtain the trajectory $\mathbf{X}_{E}\left( t\right) $. At time $t_{2}$ one
measures the position of the particle $\mathbf{x}$. Then from Fig. \ref{f1}
one gets for the platform trajectory 
\begin{equation}
\mathbf{X}\left( t_{2}\right) =\mathbf{X}_{E}\left( t_{2}\right) -\mathbf{x}.
\label{9}
\end{equation}%
The position of the particle at the time of launch in the system $\left\{
X,Y,Z\right\} $, obviously coincides with the position of the platform, $%
\mathbf{X}_{E}\left( t_{1}\right) =\mathbf{X}\left( t_{1}\right) $. To
determine the trajectory $\mathbf{X}_{E}\left( t\right) $, it is also
necessary to get the initial velocity of the particle, $\mathbf{v}_{E}\left(
t_{1}\right) $, and it depends on the velocity of the platform $\vec{V}%
\left( t_{1}\right) $, the velocity of the particle $\mathbf{v}$ in the
platform frame, the initial orientation of the platform, the rotation matrix 
$R\left( t_{1}\right) $, and the initial rate of the platform rotation $%
\mathbf{\Omega }\left( t_{1}\right) $. It is also necessary to take into
account that at the moment $t_{2}$ the orientation of the system changes ,
the rotation matrix $R\left( t_{2}\right) \not=R\left( t_{1}\right) $ and
one must rotate the vector $\mathbf{x}$ from the frame $\left\{
x,y,z\right\} $ in which it is measured to the frame $\left\{ X,Y,Z\right\} $%
. Below we will derive an obvious generalization of the formula (\ref{9})
taking into account all these factors. But here we we would like to note
that for successful navigation, restoring the position of the platform, it
is also necessary to restore the parameters%
\begin{equation}
Z=\left\{ \mathbf{V}\left( t\right) ,\mathbf{\Omega }\left( t\right)
,R\left( t\right) \right\} .  \label{9.1}
\end{equation}%
However, in order to demonstrate the power of the navigation method we are
considering here, let's assume that these parameters are known exactly. We
also assume that the particles are launched with a step $T=300$ms, and the
position of the particle is measured with accuracy [standard deviation (SD)]%
\begin{equation}
\delta _{x}=1\mu .  \label{10}
\end{equation}%
Then the time dependence of the navigation error is shown in Fig.\ref{f2}.

We assume that the accuracy of
measuring the trajectory of a classical particle (\ref{10}) is 6 orders of
magnitude worse than the accuracy of measuring (\ref{4}) the trajectory of a
quantum atomic cloud, achievable by atomic interferometry methods \cite{c1}. 
\textbf{And nevertheless, ideally, the accuracy of navigation in }$1$\textbf{%
\ hour does not exceed }$150%
{\mu}%
$\textbf{\ and practically remains the same for different trajectories.}

\begin{figure}[!t]
\includegraphics[width=8cm]{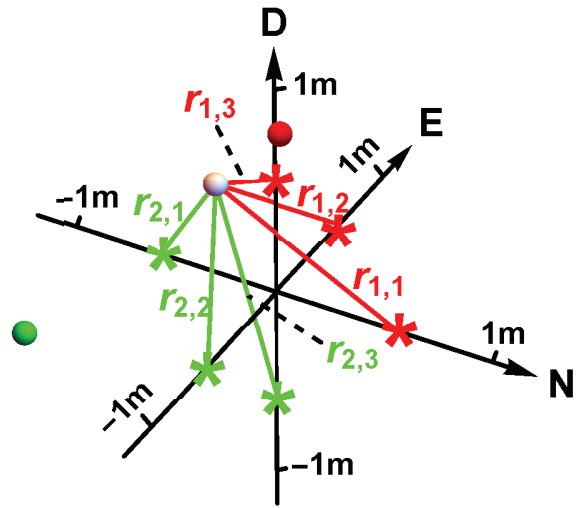}
\caption{LPS - measurement of
the particle's coordinates. From the distances to three red satellites, one
determines two positions of the particle, true white and false red.
Similarly, for green satellites, one gets true white and false green. The
true solution is obviously the same for the red and green satellites.}
\label{f3}
\end{figure}

Following \cite{c6}, in the future we will assume that the north-east-down
(NED) local frame is installed on the platform. Likewise GPS, we propose to
install in NED-frame a sensor system to measure the distance to the
particle, which we will call the local positioning system (LPS), see Fig. %
\ref{f3}.

As in GPS, we will call these sensors satellites. Our code can be used for
any satellite location. However, in the simulation, we use only satellites
located on the frame axes. When three satellites are located on the positive
semi-axes of the system, the vector connecting the particle and the
satellites do not lie in the same plane, and the coordinates of the vector $%
\mathbf{x}$ can be obtained from the distances to the satellites. The
equations for $\mathbf{x}$ have two solutions, one true, the other false. To
overcome this ambiguity, one can use an additional system of satellites
located, for example, symmetrically on negative semi-axes and choose as true
the solution that is the same for both satellite systems. During the
simulation, pseudo-random corrections were made to the exact distances using
the normal random number generator \cite{c12} with SD $\delta _{x}$.

At the launching time, we need to restore the values of the parameters (\ref%
{9.1}). Errors in restoring these parameters degrade navigation accuracy by
several orders of magnitude. These parameters can be restored using
conventional IMUs. In this case, we come to another kind of hybrid. Our code
can be included in the code of operation of such a hybrid. In this paper,
however, we will consider the possibilities of reconstructing $Z$ by
launching particles with different velocities and simultaneously measuring
their coordinates and velocities through time T.

The most significant factor affecting the accuracy of navigation is the
platform rotation. In the case of atomic navigation, it was proposed to use
the gimbal \cite{c13} to reduce the effect of rotation. In this paper, we
also assume the presence of gimbal, which we define as reference systems
that perform only noisy rotations at small angles, with zero mean value and
SD $\delta _{g}$. In the future, we will determine the quality of the
gimbals, SD $\delta _{g}$, to achieve the ultimate goal of navigation (\ref%
{6}).

An elementary loop of the navigation is the restoration of the position,
velocity, orientation and rate of rotation of the platform at the time $%
t_{2}=t_{1}+T$ from their values at the time of the particle launch $t_{1}$.
Then the restored parameters at the time $t_{2}$ are used for navigation at
the time $t_{3}=t_{2}+T$ and so on. We will call this method first order
navigation. Due to the large Doppler broadening of the spectral lines, one
could not use both the atomic interferometry \cite{c1} and the Ramsey
fringes \cite{c14} for this method. We propose installing in the NED-frame
velocimeters together with satellites. Measuring the velocity of a particle
in the platform frame will restore the velocity of the platform $V\left(
t_{2}\right) ,$ while measuring the distance difference between particles
launched at different velocities will restore the orientation of the
platform $R\left( t_{2}\right) ,$ and finally, measuring the velocity
difference of these particles will restore the rotation rate of the platform 
$\mathbf{\Omega }\left( t_{2}\right) .$

The severe requirements on the platform velocity recovery can be mitigated
in the case of the 2nd order navigation, when one measures particle's
positions at three times $\left\{ t_{1},t_{2}=t_{1}+T,t_{3}=t_{1}+2T\right\} 
$ and employs the 2nd order difference (\ref{3b}) to restore platform
position $\mathbf{X}\left( t_{3}\right) $. The reason here is that even with
an arbitrary trajectory of the platform, the difference (\ref{3b}) does not
depend on the initial velocity of the particle, if the platform does not
rotate, and the Earth's gravitational field is uniform. The weak
inhomogeneity of the field and the noisy, small rotation of the gimbal lead
to such a weak dependence of $\mathbf{p}$ on the platform velocity that
there is no need in velocimeters.

The requirements for the quality of the gimbal can also be relaxed if one
launches simultaneously 3 particles with different velocities and restores
the orientation of the gimbal from the differences in their trajectories. In
the case of the atom interferometry, iterative method of this kind
restoration was developed in \cite{c13}. Here we generalize this method to
the case of LPS-navigation (LPS-N).

We offer here 5 varieties of LPS-N:

\begin{enumerate}
\item[1.1] 1st order LPS-N, the quality of the gimbal is high enough to
neglect its rotation;

\item[1.2] 1st order LPS-N, the quality of the gimbal is sufficient to
neglect its rotation, but one does not neglect the rate of rotation, and
restores it applying data of velocimeters;

\item[1.3] 1st order LPS-N, one restores the orientation and rotation rate
of the gimbal applying satellites' and velocimeters' data;

\item[2.1] 2nd order LPS-N, the quality of the gimbal is high enough to
neglect its rotation;

\item[2.2] 2nd order LPS-N, one restores the orientation of the gimbal from
satellite data.
\end{enumerate}

The article is arranged as follows. The exact expression for the trajectory
of a particle in the field of the rotating Earth, when the potential of this
field is expanded to the 1st order gravity-gradient terms, is obtained in 
\cite{c11}. For the completeness of this study, we derive this expression in
the next section. The technique of restoring the orientation, rotation rate,
velocity, and then the position of the platform is described in Sec. \ref{s3}
for LPS-N of the 1st order and in Sec. \ref{s4} for LPS-N of the 2nd order.
The results of the simulation and their discussion were carried out in the
Sec. \ref{s5}. In the appendix Sec. \ref{A3} we describe the model of the
Earth's gravitational field accepted in our code.

\section{\label{s2}Particle trajectory}

If the source of gravity rotates at a constant velocity $\mathbf{\Omega }%
_{E} $, then the trajectory of the particle obeys the following equation of
motion 
\begin{equation}
\overset{_{\bullet \bullet }}{\mathbf{X}}_{E}=-\mathbf{\triangledown }%
U\left( \mathbf{X}_{E}\right) -\mathbf{\Omega }_{E}\times \left( \mathbf{%
\Omega }_{E}\times \mathbf{X}_{E}\right) -2\mathbf{\Omega }_{E}\times 
\mathbf{v}_{E},  \label{11}
\end{equation}%
where $\mathbf{v}_{E}=\overset{_{\bullet }}{\mathbf{X}}_{E}$ is the velocity
of the particle in the rotating Earth frame, $U\left( \mathbf{X}\right) $ is
the potential of the gravitational field. If the motion occurs in a small
vicinity of the point $\mathbf{X}_{c}$, then expanding the potential to the
second order terms,%
\begin{equation}
U\left( \mathbf{X}_{E}\right) \approx U\left( \mathbf{X}_{c}\right) -\mathbf{%
g}_{E}\left( \mathbf{X}_{c}\right) \cdot \mathbf{x}_{E}-\dfrac{1}{2}\mathbf{x%
}_{E}^{T}\underline{\Gamma }\left( \mathbf{X}_{c}\right) \mathbf{x}_{E},
\label{12}
\end{equation}%
where%
\begin{equation}
\mathbf{x}_{E}=\mathbf{X}_{E}-\mathbf{X}_{c},  \label{13}
\end{equation}%
and

\begin{subequations}
\label{13.1}
\begin{eqnarray}
\mathbf{g}_{E}\left( \mathbf{X}\right) &=&-\vec{\triangledown}U\left( 
\mathbf{X}\right) ,  \label{13.1a} \\
\underline{\Gamma }\left( \mathbf{X}\right) &=&-\mathbf{\partial }_{\mathbf{X%
}}\mathbf{\partial }_{\mathbf{X}}^{T}U\left( \mathbf{X}\right)  \label{13.1b}
\end{eqnarray}%
are the gravitational acceleration and the 1st order gravity-gradient
tensor, one comes to the equation of motion 
\end{subequations}
\begin{equation}
\overset{_{\bullet \bullet }}{\mathbf{x}}_{E}=\mathbf{g}\left( \mathbf{X}%
_{c}\right) -2\mathbf{\Omega }_{E}\times \mathbf{v}_{E}+\underline{q}\left( 
\mathbf{X}_{c}\right) \mathbf{x}_{E},  \label{14}
\end{equation}%
where 
\begin{subequations}
\label{15}
\begin{eqnarray}
\mathbf{g}\left( \mathbf{X}_{c}\right) &=&\mathbf{g}_{E}\left( \mathbf{X}%
_{c}\right) -\mathbf{\Omega }_{E}\times \left( \mathbf{\Omega }_{E}\times 
\mathbf{X}_{c}\right) ,  \label{15a} \\
\underline{q}\left( \mathbf{X}_{c}\right) &=&\underline{\Gamma }\left( 
\mathbf{X}_{c}\right) -\underline{\Omega }_{E}^{2},  \label{15b}
\end{eqnarray}%
and the tensor $\underline{\Omega }_{E}$ is given by 
\end{subequations}
\begin{equation}
\underline{\Omega }_{Eik}=-\varepsilon _{ikm}\Omega _{Em}.  \label{16}
\end{equation}%
Despite the fact that Eq. (\ref{14}) is a linear system of equations with
permanent coefficients, most of the papers used expansions of $\mathbf{x}%
_{E}\left( t\right) $ into power series over time. In the Refs. \cite%
{c14.1,c14.2}, the terms up to the 4th order were calculated, the 5th and
6th order terms were obtained in Ref. \cite{c14.3}, the expansion to the 7th
order was obtained in the article \cite{c14.4}. The exact solution of the
equation (\ref{14}) is obtained in the patent \cite{c11}. Here we derive
exact solutions for arbitrarily large values of the tensor $\underline{q}%
\left( \mathbf{X}_{c}\right) $ and the rate of rotation of the gravity field
source, $\mathbf{\Omega }_{E}$. In this case, we are using the method
proposed in \cite{c15}. Consider first the homogeneous equation $\left[ 
\mathbf{g}\left( \mathbf{X}_{c}\right) =0\right] $. Its solution is $\mathbf{%
x}_{E}\left( t\right) =\mathbf{x}\left( \omega \right) e^{-i\omega t}$,
where $\omega $ and $\mathbf{x}\left( \omega \right) $ are the eigenvalue
and eigenvector of the matrix%
\begin{equation}
\underline{A}\left( \omega \right) =\underline{q}\left( \mathbf{X}%
_{c}\right) +\omega ^{2}I+2i\omega \underline{\Omega }_{E},  \label{18}
\end{equation}%
where $I$ is the unity matrix. One notices that owing the symmetry of the
tensor $\underline{q}\left( \mathbf{X}_{c}\right) ,$ the determinant $%
\left\vert \underline{A}\left( \omega \right) \right\vert $ is an even
function $\omega $, $\left\vert \underline{A}\left( \omega \right)
\right\vert =\left\vert \underline{A}\left( -\omega \right) \right\vert $,
and, therefore, the characteristic.equation%
\begin{equation}
\left\vert \underline{A}\left( \omega \right) \right\vert =0,  \label{17}
\end{equation}%
is reduced to the 3rd order equation for $\omega ^{2}$%
\begin{equation}
\omega ^{6}+a_{2}\omega ^{4}+a_{1}\omega ^{2}+a_{0}=0,  \label{19}
\end{equation}%
where 
\begin{subequations}
\label{20}
\begin{gather}
a_{0}=\left\vert q\left( \mathbf{X}_{c}\right) \right\vert ,  \label{20a} \\
a_{1}=\dfrac{1}{2}\left\{ Tr^{2}\left[ \underline{q}\left( \mathbf{X}%
_{c}\right) \right] -Tr\left[ \underline{q}^{2}\left( \mathbf{X}_{c}\right) %
\right] \right\} -4\mathbf{\Omega }_{E}^{T}\underline{q}\Omega _{En},
\label{20b} \\
a_{2}=Tr\left[ \underline{q\left( \mathbf{X}_{c}\right) }\right] -4\Omega
_{E}^{2}.  \label{20c}
\end{gather}%
Eq. (\ref{19}) has 6 roots 
\end{subequations}
\begin{subequations}
\label{21}
\begin{eqnarray}
\omega _{i} &=&\sqrt{z_{i}-\dfrac{a_{2}}{3}},~\text{for }i=1,2,3  \label{21a}
\\
\left\{ \omega _{4},\omega _{5},\omega _{6}\right\} &=&-\left\{ \omega
_{1},\omega _{2},\omega _{3}\right\} ,  \label{21b}
\end{eqnarray}%
where $z_{i}$ is the root of the equation $z^{3}+pz+s=0$ and 
\end{subequations}
\begin{subequations}
\label{22}
\begin{eqnarray}
p &=&a_{1}-\dfrac{a_{2}^{2}}{3},  \label{22a} \\
s &=&a_{0}-\dfrac{a_{2}a_{1}}{3}+\dfrac{2a_{2}^{3}}{27}.  \label{22b}
\end{eqnarray}%
The eigenvectors are given by 
\end{subequations}
\begin{equation}
\mathbf{x}\left( \omega \right) =\left( 
\begin{array}{c}
A_{22}\left( \omega \right) A_{33}\left( \omega \right) -A_{23}\left( \omega
\right) A_{32}\left( \omega \right) \\ 
A_{23}\left( \omega \right) A_{31}\left( \omega \right) -A_{21}\left( \omega
\right) A_{33}\left( \omega \right) \\ 
A_{21}\left( \omega \right) A_{32}\left( \omega \right) -A_{22}\left( \omega
\right) A_{31}\left( \omega \right)%
\end{array}%
\right) .  \label{23}
\end{equation}%
Let's return now to the equation (\ref{14}). We introduce a 6-dimensional
vector%
\begin{equation}
\mathbf{y}=\left( 
\begin{array}{c}
\mathbf{x}_{E} \\ 
\mathbf{v}_{E}%
\end{array}%
\right) ,  \label{24}
\end{equation}%
which obeys the equation%
\begin{equation}
\mathbf{\dot{y}}=\left( 
\begin{array}{cc}
0 & I \\ 
\underline{q} & -2\underline{\Omega }_{E}%
\end{array}%
\right) \mathbf{y}+\left( 
\begin{array}{c}
0 \\ 
\mathbf{g}%
\end{array}%
\right) .  \label{25}
\end{equation}%
One gets that solution is given by%
\begin{eqnarray}
\mathbf{y}\left( t\right) &=&\underline{\Phi }\left( t\right) \mathbf{y}%
\left( 0\right) +\int_{0}^{t}dt^{\prime }\underline{\Phi }\left( t-t^{\prime
}\right) \left( 
\begin{array}{c}
0 \\ 
\mathbf{g}%
\end{array}%
\right) ,  \label{26a} \\
\underline{\Phi }\left( t\right) &=&\underline{Y}\left( t\right) \underline{Y%
}^{-1}\left( 0\right) ,  \label{26b}
\end{eqnarray}%
where $6\times 6$ matrix $\underline{Y}\left( t\right) $ is
composed of eigenvectors,%
\begin{equation}
\underline{Y}\left( t\right) =\left\{ \mathbf{y}_{1}e^{-i\omega
_{1}t},\ldots ,\mathbf{y}_{6}e^{-i\omega _{6}t}\right\} .  \label{27}
\end{equation}%
Since%
\begin{equation}
\mathbf{y}_{m}=\left( 
\begin{array}{c}
\mathbf{x}_{m} \\ 
-i\omega _{m}\mathbf{x}_{m}%
\end{array}%
\right) ,  \label{28}
\end{equation}%
one can represent the matrix (\ref{27}) as%
\begin{equation}
\underline{Y}\left( t\right) =\left( 
\begin{array}{cc}
\underline{x}_{+}e^{-i\underline{\tilde{\omega}}t} & \underline{x}_{-}e^{i%
\underline{\tilde{\omega}}t} \\ 
-i\underline{x}_{+}\underline{\tilde{\omega}}e^{-i\underline{\tilde{\omega}}%
t} & i\underline{x}_{-}\underline{\tilde{\omega}}e^{i\underline{\tilde{\omega%
}}t}%
\end{array}%
\right) ,  \label{29}
\end{equation}%
where $3\times 3-$matrices \underline{$\underline{x}$}$_{\pm }$ and $\omega $
are composed of eigenvectors (\ref{23}) and eigenfrequencies 
\begin{subequations}
\label{30}
\begin{eqnarray}
\underline{x}_{\pm } &\equiv &\left\{ \mathbf{x}\left( \pm \omega
_{1}\right) ,\mathbf{x}\left( \pm \omega _{2}\right) ,\mathbf{x}\left( \pm
\omega _{3}\right) \right\} ,  \label{30a} \\
\underline{\tilde{\omega}} &=&diag\left( \omega _{1},\omega _{2},\omega
_{3}\right) .  \label{30b}
\end{eqnarray}%
To inverse the matrix $\underline{Y}\left( 0\right) $, one can use the
method developed in \cite{c16}. It is convenient to represent the $6\times
6- $matrices $\underline{\Phi }\left( t\right) $ and $\underline{Y}%
^{-1}\left( 0\right) $ as $2\times 2-$matrices 
\end{subequations}
\begin{subequations}
\label{31}
\begin{eqnarray}
\underline{\Phi }\left( t\right) &=&\left( 
\begin{array}{cc}
\underline{x}_{x}\left( t\right) & \underline{x}_{v}\left( t\right) \\ 
\underline{\dot{x}}_{x}\left( t\right) & \underline{\dot{x}}_{v}\left(
t\right)%
\end{array}%
\right) ,  \label{31a} \\
\underline{Y}^{-1}\left( 0\right) &=&\left( 
\begin{array}{cc}
b_{11} & b_{12} \\ 
b_{21} & b_{22}%
\end{array}%
\right) ,  \label{31b}
\end{eqnarray}%
in which the matrix elements are $3\times 3-$matrices. Then one gets for $%
b_{mn}$%
\end{subequations}
\begin{subequations}
\label{32}
\begin{eqnarray}
\underline{x}_{+}b_{11}+\underline{x}_{-}b_{21} &=&I,  \label{32a} \\
\underline{x}_{+}\underline{\tilde{\omega}}b_{11}-\underline{x}_{-}%
\underline{\tilde{\omega}}b_{21} &=&0,  \label{32b} \\
\underline{x}_{+}\underline{\tilde{\omega}}b_{12}-\underline{x}_{-}%
\underline{\tilde{\omega}}b_{22} &=&i,  \label{32c} \\
\underline{x}_{+}b_{12}+\underline{x}_{-}b_{22} &=&0.  \label{32d}
\end{eqnarray}%
Solving the Eqs. (\ref{32}) and substituting their solutions consequently in
the Eqs. (\ref{26b}) and (\ref{26a}), after integrating into the Eq. (\ref%
{26a}), one arrives at the next exact solution of the Eq. (\ref{14}) 
\end{subequations}
\begin{subequations}
\label{33}
\begin{align}
\mathbf{x}_{E}\left( t\right) & =\underline{x}_{x}\left( t\right) \mathbf{x}%
_{E}\left( 0\right) +\underline{x}_{v}\left( t\right) \mathbf{v}_{E}\left(
0\right) +\underline{x}_{g}\left( t\right) \mathbf{g}\left( \mathbf{X}%
_{c}\right) ,  \label{33a} \\
\underline{x}_{x}\left( t\right) & =\left( \underline{x}_{+}e^{-i\underline{%
\tilde{\omega}}t}+\underline{x}_{-}e^{i\underline{\tilde{\omega}}t}%
\underline{\tilde{\omega}}^{-1}\underline{x}_{-}^{-1}\underline{x}_{+}%
\underline{\tilde{\omega}}\right) \underline{x}_{i},  \label{33b} \\
\underline{x}_{v}\left( t\right) & =i\left( \underline{x}_{+}e^{-i\underline{%
\tilde{\omega}}t}\underline{x}_{+}^{-1}\underline{x}_{-}-\underline{x}%
_{-}e^{i\underline{\tilde{\omega}}t}\right) \underline{v}_{i}  \label{33c} \\
\underline{x}_{g}\left( t\right) & =\left( \underline{x}_{+}\dfrac{e^{-i%
\underline{\tilde{\omega}}t}-1}{\underline{\tilde{\omega}}}\underline{x}%
_{+}^{-1}\underline{x}_{-}+\underline{x}_{-}\dfrac{e^{i\underline{\tilde{%
\omega}}t}-1}{\underline{\tilde{\omega}}}\right) \underline{v}_{i},
\label{33d}
\end{align}%
where 
\end{subequations}
\begin{subequations}
\label{34}
\begin{eqnarray}
\underline{x}_{i} &=&\left( x_{+}+x_{-}\tilde{\omega}^{-1}x_{-}^{-1}x_{+}%
\tilde{\omega}\right) ^{-1},  \label{34a} \\
\underline{v}_{i} &=&\left( x_{-}\tilde{\omega}+x_{+}\tilde{\omega}%
x_{+}^{-1}x_{-}\right) ^{-1}.  \label{34b}
\end{eqnarray}

\section{\label{s3}1st order LPS-N}

Suppose that in the platform frame, one launches a particle at the time $t=0$
from the point $\mathbf{x}\left( 0\right) $ at the velocity $\mathbf{v}%
\left( 0\right) $. After the interrogation time $T$, one measures the
position and velocity of the particle $\left\{ \mathbf{x}\left( T\right) ,%
\mathbf{v}\left( T\right) \right\} $ using the local system of satellites
and velocimeters and applying Eqs. (\ref{a1}, \ref{b3}). The navigation
purpose is to restore the position, orientation, velocity, and rotation rate
of the platform, $\left\{ \mathbf{X}\left( T\right) ,R\left( T\right) ,%
\mathbf{V}\left( T\right) ,\mathbf{\Omega }\left( T\right) \right\} ,$ from
these measurements.

After the launch, the particles are decoupled from the platform and move
into the Earth gravitational and inertial fields.. Its position in respect
to the Earth origin is given by 
\end{subequations}
\begin{equation}
\mathbf{X}_{E}\left( t\right) =\mathbf{X}_{c}+\mathbf{x}_{E}\left( t\right) ,
\label{35}
\end{equation}%
where $\mathbf{x}_{E}\left( t\right) $ one obtains from Eq. (\ref{33a}). If
the origin of the platform frame moves along the trajectory $\mathbf{X}%
\left( T\right) $, then%
\begin{equation}
\mathbf{x}\left( T\right) =R\left( T\right) \left[ \mathbf{x}_{E}\left(
T\right) +\mathbf{X}_{c}-\mathbf{X}\left( T\right) \right] .  \label{36}
\end{equation}%
Since the derivative of the rotation matrix is determined by the rotation
rate, $\dot{R}\left( T\right) =-\underline{\Omega }\left( T\right) R\left(
T\right) ,$ one obtains for the particle velocity%
\begin{equation}
\mathbf{v}\left( T\right) =-\mathbf{\Omega }\left( T\right) \times \mathbf{x}%
\left( T\right) +R\left( T\right) \left[ \mathbf{v}_{E}\left( T\right) -%
\mathbf{V}\left( T\right) \right] .  \label{37}
\end{equation}%
The particle is detached from the platform only after launch, at $T>0$. If $%
T=0$, one gets from Eqs. (\ref{36}, \ref{37}) 
\begin{subequations}
\label{38}
\begin{eqnarray}
\mathbf{x}_{E}\left( 0\right) &=&\mathbf{X}\left( 0\right) -\mathbf{X}%
_{c}+R^{T}\left( 0\right) \mathbf{x}\left( 0\right) ,  \label{38a} \\
\mathbf{v}_{E}\left( 0\right) &=&\mathbf{V}\left( 0\right) +R^{T}\left(
0\right) \left[ \mathbf{v}\left( 0\right) +\mathbf{\Omega }\times \mathbf{x}%
\left( 0\right) \right] .  \label{38b}
\end{eqnarray}%
Then one gets 
\end{subequations}
\begin{subequations}
\label{39}
\begin{eqnarray}
\mathbf{x}\left( T\right) &=&R\left( T\right) \left\{ \underline{x}%
_{x}\left( T\right) \left[ \mathbf{X}\left( 0\right) -\mathbf{X}%
_{c}+R^{T}\left( 0\right) \mathbf{x}\left( 0\right) \right] \right.  \notag
\\
&&+\underline{x}_{v}\left( T\right) \left[ \mathbf{V}\left( 0\right)
+R^{T}\left( 0\right) \left( \mathbf{v}\left( 0\right) +\mathbf{\Omega }%
\left( 0\right) \times \mathbf{x}\left( 0\right) \right) \right]  \notag \\
&&\left. +\underline{x}_{g}\left( T\right) \mathbf{g}\left( \mathbf{X}%
_{c}\right) +\mathbf{X}_{c}-\mathbf{X}\left( T\right) \right\} ;  \label{39a}
\\
\mathbf{v}\left( T\right) &=&-\mathbf{\Omega }\left( T\right) \times \mathbf{%
x}\left( T\right) +R\left( T\right)  \notag \\
&&\times \left\{ \underline{\dot{x}}_{x}\left( T\right) \left[ \mathbf{X}%
\left( 0\right) -\mathbf{X}_{c}+R^{T}\left( 0\right) \mathbf{x}\left(
0\right) \right] \right.  \notag \\
&&+\underline{\dot{x}}_{v}\left( T\right) \left[ \mathbf{V}\left( 0\right)
+R^{T}\left( 0\right) \left( \mathbf{v}\left( 0\right) +\mathbf{\Omega }%
\left( 0\right) \times \mathbf{x}\left( 0\right) \right) \right]  \notag \\
&&\left. +\underline{\dot{x}}_{g}\left( T\right) \mathbf{g}\left( \mathbf{X}%
_{c}\right) -\mathbf{V}\left( T\right) \right\} .  \label{39b}
\end{eqnarray}%
The equation (\ref{39a}) gives the exact trajectory of the particle in the
platform frame, which can, for example, be used to exact calculation of the
AI phase, instead of the approximate calculations in the articles \cite%
{c14.1,c14.2,c14.3,c14.4}. Like in the previous studies \cite{c11,c13}, here
we are going to use Eq. (\ref{39}) to restore the position and velocity of
the platform. For them, one has 
\end{subequations}
\begin{gather}
\mathbf{X}\left( T\right) =\mathbf{X}_{c}+\underline{x}_{x}\left( T\right) %
\left[ \mathbf{X}\left( 0\right) -\mathbf{X}_{c}+R^{T}\left( 0\right) 
\mathbf{x}\left( 0\right) \right]  \notag \\
+\underline{x}_{v}\left( T\right) \left[ \mathbf{V}\left( 0\right)
+R^{T}\left( 0\right) \left( \mathbf{v}\left( 0\right) +\mathbf{\Omega }%
\left( 0\right) \times \mathbf{x}\left( 0\right) \right) \right]  \notag \\
+\underline{x}_{g}\left( T\right) \mathbf{g}\left( \mathbf{X}_{c}\right)
-R^{T}\left( T\right) \mathbf{x}\left( T\right) ,  \label{40a} \\
\mathbf{V}\left( T\right) =\underline{\dot{x}}_{x}\left( T\right) \left[ 
\mathbf{X}\left( 0\right) -\mathbf{X}_{c}+R^{T}\left( 0\right) \mathbf{x}%
\left( 0\right) \right]  \notag \\
+\underline{\dot{x}}_{v}\left( t\right) \left[ \mathbf{V}\left( 0\right)
+R^{T}\left( 0\right) \left( \mathbf{v}\left( 0\right) +\mathbf{\Omega }%
\times \mathbf{x}\left( 0\right) \right) \right]  \notag \\
+\underline{\dot{x}}_{g}\left( T\right) \mathbf{g}\left( \mathbf{X}%
_{c}\right) -R^{T}\left( T\right) \left[ \mathbf{v}\left( T\right) +\mathbf{%
\Omega }\left( T\right) \times \mathbf{x}\left( T\right) \right] .
\label{40b}
\end{gather}

\subsection{\label{s3.1}Rotation rate restore}

One sees that in order to obtain the position and velocity of the platform,
it is necessary to initially restore the rotation rate $\mathbf{\Omega }%
\left( T\right) $ and the rotation matrix $R\left( T\right) $. For these
restorations, one can use a differential technique. If from the same point
one launches 3 particles with velocities $\mathbf{v},$ $\mathbf{v}^{\prime }$
and $\mathbf{v}^{\prime \prime }$, then using Eq.. (\ref{39b}) to get the
velocity differences at the time $T$%
\begin{subequations}
\label{40.1}
\begin{eqnarray}
\delta \mathbf{v}_{1}\left( T\right) &=&\mathbf{v}\left( T\right) -\mathbf{v}%
^{\prime }\left( T\right) ,  \label{40.1a} \\
\delta \mathbf{v}_{2}\left( T\right) &=&\mathbf{v}\left( T\right) -\mathbf{v}%
^{\prime \prime }\left( T\right) .  \label{40.1b}
\end{eqnarray}%
one arrives to the following equations for $\mathbf{\Omega }\left( T\right) $%
\end{subequations}
\begin{equation}
\mathbf{a}_{i}\times \mathbf{\Omega }\left( T\right) =\mathbf{b}_{i},
\label{41}
\end{equation}%
where 
\begin{subequations}
\label{42}
\begin{eqnarray}
\mathbf{a}_{i} &\equiv &\delta \mathbf{x}_{i}\left( T\right) ,  \label{42a}
\\
\mathbf{b}_{i} &\equiv &\delta \mathbf{v}_{i}\left( T\right) -R\left(
T\right) \dot{x}_{v}\left( T\right) R^{T}\left( 0\right) \delta \mathbf{v}%
_{i},  \label{42b} \\
\delta \mathbf{v}_{i} &\equiv &\delta \mathbf{v}_{i}\left( 0\right) .
\label{42c}
\end{eqnarray}%
One sees that one velocity difference,for example $\delta \mathbf{v}_{1},$
is not sufficient, it allows us to restore only the component of the
rotation rate perpendicular to $\mathbf{a}_{1}.$ That is why one starts 3
particles and considers 2 velocity differences. Decomposing $\mathbf{\Omega }%
\left( T\right) $ as 
\end{subequations}
\begin{equation}
\mathbf{\Omega }\left( T\right) =\mathbf{\Omega }_{\perp }\left( T\right)
+\Omega _{\parallel }\left( T\right) \mathbf{a}_{1},  \label{43}
\end{equation}%
one gets from Eq. (\ref{41}) for $i=1$%
\begin{equation}
\mathbf{\Omega }_{\perp }\left( T\right) =\dfrac{\mathbf{b}_{1}\times 
\mathbf{a}_{1}}{a_{1}^{2}}  \label{44}
\end{equation}%
Substituting Eq.. (\ref{44}) in Eq. (\ref{43}) and then in Eq. (\ref{41})
for i-2, one obtains the equation for $\Omega _{\parallel }\left( T\right) .$
After solving it, one arrives to the following result 
\begin{subequations}
\label{45}
\begin{gather}
\mathbf{\Omega }\left( t\right) =\mathbf{f}\left( \mathbf{a}_{i},\mathbf{b}%
_{i}\right) ,  \label{45a} \\
\mathbf{f}\left( \mathbf{a}_{i},\mathbf{b}_{i}\right) =\dfrac{\mathbf{b}%
_{1}\times \mathbf{a}_{1}}{a_{1}^{2}}  \notag \\
+\dfrac{\mathbf{a}_{1}\cdot \left\{ \left[ \mathbf{b}_{2}a_{1}^{2}-\mathbf{b}%
_{1}\left( \mathbf{a}_{1}\cdot \mathbf{a}_{2}\right) \right] \times \mathbf{a%
}_{2}\right\} }{\left( \mathbf{a}_{2}\times \mathbf{a}_{1}\right)
^{2}a_{1}^{2}}\mathbf{a}_{1}.  \label{45b}
\end{gather}

\subsection{\label{s3.2}Rotation matrix restore}

Here, also, one should use the differential technique \cite{c11,c13}. For
the same 3 particles with launch velocities $\mathbf{v},$ $\mathbf{v}%
^{\prime }$ and $\mathbf{v}^{\prime \prime }$ instead of the velocities'
difference (\ref{40.1}), one now measures the positions' differences 
\end{subequations}
\begin{subequations}
\label{46}
\begin{eqnarray}
\delta \mathbf{x}_{1}\left( T\right) &=&\mathbf{x}\left( T\right) -\mathbf{x}%
^{\prime }\left( T\right) ,  \label{46a} \\
\delta \mathbf{x}_{2}\left( T\right) &=&\mathbf{x}\left( T\right) -\mathbf{x}%
^{\prime \prime }\left( T\right) ,  \label{46b}
\end{eqnarray}%
which depend only on the rotation matrices, 
\end{subequations}
\begin{subequations}
\label{47}
\begin{eqnarray}
\delta \mathbf{x}_{i}\left( T\right) &=&r\mathbf{a}_{i},  \label{47a} \\
r &=&R\left( T\right) R^{T}\left( 0\right) ,  \label{47b} \\
\mathbf{a}_{i} &=&R\left( 0\right) \underline{x}_{v}\left( T\right)
R^{T}\left( 0\right) \delta \mathbf{v}_{i},  \label{47c}
\end{eqnarray}%
where $\delta \mathbf{v}_{i}$ is defined in Eq. (\ref{42c}). The matrix $r$
in the Rodriguez representation \cite{c16.1} is given by 
\end{subequations}
\begin{equation}
r=\cos \delta +\dfrac{\left( 1-\cos \delta \right) }{\delta ^{2}}\mathbf{%
\delta \delta }^{T}-\dfrac{\sin \delta }{\delta }\underline{\delta },
\label{49}
\end{equation}%
where the vector $\mathbf{\delta }$ contains 3 parameters that define the
rotation. In the Ref. \cite{c13}, an iterative method was proposed for
calculating the angle of rotation $\mathbf{\delta }$ from the phase
differences of $6$ AIs, in which the atoms were launched at different
speeds, and the wave vectors did not lie in the same plane. Here we adapt
this method for the case of LPS.

In the case of rotation at a small angle, $\delta \ll 1$, when $r\approx 1-$%
\underline{$\delta $}, and matrix $\delta $ defined as%
\begin{equation}
\underline{\delta }_{ik}=-\varepsilon _{ikm}\delta _{m},  \label{49.1}
\end{equation}%
one gets that%
\begin{equation}
\mathbf{a}_{i}\times \mathbf{\delta }=\mathbf{b}_{i},  \label{50}
\end{equation}%
where%
\begin{equation}
\mathbf{b}_{i}=\delta \mathbf{x}_{i}\left( T\right) -\mathbf{a}_{i},
\label{51}
\end{equation}%
and $\left\vert b_{i}\right\vert \ll \left\vert a_{i}\right\vert $. This
equation coincides with Eq. (\ref{41}), and hence,%
\begin{equation}
\mathbf{\delta }=\mathbf{f}\left( \mathbf{a}_{i},\mathbf{b}_{i}\right) ,
\label{52}
\end{equation}%
where $\mathbf{f}$ is the vector function defined in Eq. (\ref{45b}).

Let us now turn to the general case. Then, from Eq. (\ref{47a}), one gets%
\begin{equation}
\sin \delta \mathbf{a}_{i}\times \mathbf{n}+\left( 1-\cos \delta \right) %
\left[ \mathbf{n}\left( \mathbf{na}_{i}\right) -\mathbf{a}_{i}\right] -%
\mathbf{b}_{i}=0,  \label{53}
\end{equation}%
where $\mathbf{n}=\mathbf{\delta }/\delta $. One can introduce the parameter%
\begin{equation}
\sigma =\left\vert \mathbf{f}\left( \mathbf{a}_{i},\mathbf{b}_{i}\right)
\right\vert  \label{54}
\end{equation}%
and expand the solutions of Eq. (\ref{53}) in the series 
\begin{subequations}
\label{55}
\begin{eqnarray}
\delta &=&\dsum_{s=1}^{\infty }\delta ^{\left( s\right) }\sigma ^{s},
\label{55a} \\
\mathbf{n} &=&\dsum_{s=0}^{\infty }\mathbf{n}^{\left( s\right) }\sigma ^{s}
\label{55b}
\end{eqnarray}%
The lower order terms in these series are equal 
\end{subequations}
\begin{subequations}
\label{56}
\begin{eqnarray}
\delta ^{\left( 1\right) } &=&1,  \label{56a} \\
\mathbf{n}^{\left( 0\right) } &=&\mathbf{f}\left( \mathbf{a}_{i},\mathbf{b}%
_{i}\right) /\sigma  \label{56b}
\end{eqnarray}%
In Appendix \ref{A2} , we have derived the recurrence relations for the
coefficients $\left\{ \mathbf{n}^{\left( s-1\right) },\delta ^{\left(
s\right) }\right\} $.

\section{\label{s4}2nd order LPS-N}

In this case, one restores the trajectory and orientation of the platform $%
\left\{ \mathbf{X}\left( 2T\right) ,R\left( 2T\right) \right\} $ from their
previous values at moments $t=0$ and $t=T$, using a second-order difference 
\end{subequations}
\begin{equation}
\mathbf{p}=\mathbf{x}\left( 2T\right) -2x\left( T\right) +\mathbf{x}\left(
0\right) ,  \label{57}
\end{equation}%
in which the particle positions in the platform frame, $\mathbf{x}\left(
T\right) $ and $x\left( 2T\right) $ are measured using satellites.

An important advantage of such navigation is a tremendous decrease of the
navigation sensitivity to the initial velocity of the platform. For
successful navigation, it is sufficient to put \cite{c11,c13}%
\begin{equation}
\mathbf{V}\left( 0\right) =\dfrac{\mathbf{X}\left( T\right) -\mathbf{X}%
\left( -T\right) }{2T},  \label{58}
\end{equation}%
where, for a given loop of the navigation cycle, $t\in \left[ 0,2T\right] $,
the positions of the platform $\mathbf{X}\left( -T\right) $ and $\mathbf{X}%
\left( T\right) $ are restored in the previous loops.

If all the particles are launched from the center of the platform $\left[ 
\mathbf{x}\left( 0\right) =0\right] ,$ then the lever=arm term, $\mathbf{%
\Omega }\left( 0\right) \times \mathbf{x}\left( 0\right) ,$ disappears, and
simultaneously the need to restore the angular velocity of the platform
rotation $\mathbf{\Omega }\left( 0\right) $ disappears also. Let's also
assume for simplicity that $\mathbf{X}_{c}=\mathbf{X}\left( 0\right) $. Then%
\begin{eqnarray}
\mathbf{p} &=&\left[ R\left( 2T\right) \underline{x}_{v}\left( 2T\right)
-2R\left( T\right) \underline{x}_{v}\left( T\right) \right]  \notag \\
&&\times \left[ \mathbf{V}\left( 0\right) +R^{T}\left( 0\right) \mathbf{v}%
\left( 0\right) \right]  \notag \\
&&+\left[ R\left( 2T\right) \underline{x}_{g}\left( 2T\right) -2R\left(
T\right) \underline{x}_{g}\left( T\right) \right] \mathbf{g}\left( \mathbf{X}%
\left( 0\right) \right)  \notag \\
&&-R\left( 2T\right) \left[ \mathbf{X}\left( 2T\right) -\mathbf{X}\left(
0\right) \right] +2R\left( T\right) \left[ \mathbf{X}\left( T\right) -%
\mathbf{X}\left( 0\right) \right] .  \label{59}
\end{eqnarray}%
In the absence of rotation $\left[ R\left( t\right) =I\right] $ and for a
uniform, non-rotating gravitational field%
\begin{equation}
\underline{x}_{v}\left( t\right) =t;  \label{60}
\end{equation}%
the first term in Eq. (\ref{59}) disappears and the difference $\mathbf{p}$
does not depend on the initial velocity of the platform. The simulation
showed that despite the weak inhomogeneity of the field, the presence of
inertial forces due to the rotation of the Earth and the small rotations of
the gimbal, the first term remains so small that the error in restoring the
velocity (\ref{58}) has a much smaller impact on the accuracy of navigation
than other factors, and one may not take it into account. Then the
coordinate of the platform $\mathbf{X}\left( 2T\right) $ is restored as%
\begin{eqnarray}
\mathbf{X}\left( 2T\right) &=&R^{T}\left( 2T\right) \left\{ -\mathbf{p}+%
\left[ R\left( 2T\right) \underline{x}_{v}\left( 2T\right) -2R\left(
T\right) \underline{x}_{v}\left( T\right) \right] \right.  \notag \\
&&\times \left[ \mathbf{V}\left( 0\right) +R^{T}\left( 0\right) \mathbf{v}%
\left( 0\right) \right]  \notag \\
&&+\left[ R\left( 2T\right) \underline{x}_{g}\left( 2T\right) -2R\left(
T\right) \underline{x}_{g}\left( T\right) \right] \mathbf{g}\left( \mathbf{X}%
\left( 0\right) \right)  \notag \\
&&\left. +2R\left( T\right) \mathbf{X}\left( T\right) +\left[ R\left(
2T\right) -2R\left( T\right) \right] \mathbf{X}\left( 0\right) \right\} .
\label{61}
\end{eqnarray}%
If one neglects the rotation of the gimbal, assuming $R\left( 2T\right)
=R\left( T\right) =R\left( 0\right) $, then no further steps are required.
Otherwise, like in Sec. \ref{s3.2}, one launches simultaneously from the
origin 3 particles with velocities $\mathbf{v}\left( 0\right) =\mathbf{v},$ $%
\mathbf{v}^{\prime },$ and $\mathbf{v}^{\prime \prime }$ and measures the
double differences 
\begin{subequations}
\label{62}
\begin{eqnarray}
\delta \mathbf{p}_{1}\left( T\right) &=&\mathbf{p}\left( T\right) -\mathbf{p}%
^{\prime }\left( T\right) ,  \label{62a} \\
\delta \mathbf{p}_{2}\left( T\right) &=&\mathbf{p}\left( T\right) -\mathbf{p}%
^{\prime \prime }\left( T\right) ,  \label{62b}
\end{eqnarray}%
which lead to the equations for the rotation matrix $R\left( 2T\right) $. In
this case, one comes to the equation 
\end{subequations}
\begin{equation}
\delta \mathbf{p}_{i}+2R\left( T\right) \underline{x}_{v}\left( T\right)
R^{T}\left( 0\right) \delta \mathbf{v}_{i}=ra_{i}  \label{63}
\end{equation}%
where 
\begin{subequations}
\label{64}
\begin{eqnarray}
r &=&R\left( 2T\right) R^{T}\left( 0\right) ,  \label{64a} \\
\mathbf{a}_{i} &=&R\left( 0\right) \underline{x}_{v}\left( 2T\right)
R^{T}\left( 0\right) \delta \mathbf{v}_{i},  \label{64b}
\end{eqnarray}%
where $\delta \mathbf{v}_{i}$ is defined in Eq. (\ref{42c}). After that one
follows the method of solving Eq. (\ref{47a}), described above and in
appendix \ref{A2}, with the only difference that now $\mathbf{a}_{i}$ is
given by Eq. (\ref{64b}) and 
\end{subequations}
\begin{equation}
\mathbf{b}_{i}=\delta \mathbf{p}_{i}+2R\left( T\right) \underline{x}%
_{v}\left( T\right) R^{T}\left( 0\right) \delta \mathbf{v}_{i}-\mathbf{a}%
_{i}.  \label{65}
\end{equation}

\section{\label{s5}Simulation results}

\begin{figure*}[t]
\includegraphics[width=1.9\columnwidth]{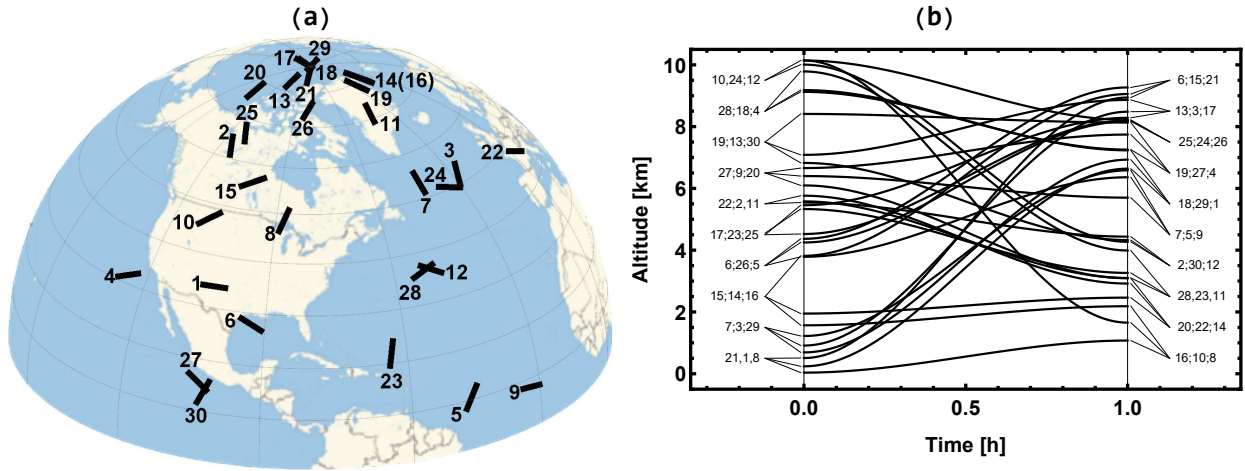}
\caption{Platform trajectories. Platform
moves along each trajectory during the time $t_{n}=1$hour. The trajectories
are chosen so that the platform velocity $V\left( 0\right) =V\left(
t_{n}\right) =0\ $and it reaches a maximum at $t=t_{n}/2$. Only those
trajectories were selected for which $V\left( t_{n}/2\right) \in \left[ 200%
\text{m/s,}240\text{m/s}\right] $. (a) Trajectories latitudes and
longitudes. (b) Time dependences of the platform trajectories' altitudes.
(The trajectories were randomly generated in the western part of the Earth
northern hemisphere for the sole purpose that they were visible in Fig. 
\protect\ref{f4}a) \label{f4}  }
\end{figure*}

We perform testing the LPS-N on an ensemble of 30 randomly generated
platform trajectories, shown in Fig. \ref{f4}.

Regarding the orientation, we assume that the North-East-Down (NED) frame
has been installed initially on the platform, i.e. the initial gimbal
rotation matrix is given by%
\begin{equation}
R\left( 0\right) =\left( 
\begin{array}{ccc}
-\sin \phi \cos \lambda & -\sin \phi \sin \lambda & \cos \phi \\ 
-\sin \lambda \cos \lambda & \cos \lambda & 0 \\ 
-\cos \phi \cos \lambda & -\cos \phi \sin \lambda & -\sin \phi%
\end{array}%
\right) ,  \label{66}
\end{equation}%
where $\phi $ and $\lambda $ are initial latitudes and longitude of the
platform. Later on gimbal commits only random noisy rotations at small
angles of the order of gimbal stabilization level $\delta _{g}$, in the
frequency band $f_{g}=100$Hz.

Our navigation code consists of the following points

\begin{enumerate}
\item One calculates the exact values of the particle's coordinate and
velocity using the Eqs. (\ref{39}, \ref{33b} - \ref{33d}) and exact
expressions for the trajectory and rotation matrix of the platform and the
Earth's gravitational field (see appendix \ref{A3} below).

\item Coordinates and velocities provide us accurate "readings" of
satellites and velocimeters. One exploits the pseudo-random normal number
generators \cite{c12} with SD $\delta _{x}$. and $\delta _{v}$ to make
corrections to the sensors' readings

\item The sensor data are the input parameters for restoring the coordinates
and velocities in the platform frame using Eqs. (\ref{a1}, \ref{b3}).

\item The orientation of the gimbal is either assumed unchanged (for methods
1.1, 1.2, 2.1), or restored (for methods 1.3 and 2.2). In the last case, one
solves Eq. (\ref{53}) using the iterative method (see appendix \ref{A2}).

\item If necessary, one restores the instantaneous rotation rate of the
gimbal $\mathbf{\Omega }\left( T\right) $ [according to the formula (\ref%
{45a})].

\item Finally, after restoring the velocity of the platform [see Eqs. (\ref%
{40b}) or (\ref{58})], one obtains the desired coordinates of the platform
from the Eqs. (\ref{40a}) or (\ref{61})
\end{enumerate}

With the exception of the first, in the each loop of the navigation cycle,
one uses the initial position, speed, orientation, and rotation speed of the
platform restored in the previous loop. As a result, the navigation error $%
\Delta X$ accumulates and increases on 7 orders of magnitude, from $1%
{\mu}%
$ to $10$m at the end of the cycle.

The accuracy of navigation increases with more accurate measurement of the
position of the particles in the platform frame. To achieve it, we propose
using not one set of satellites and velocimeters, but ensembles of these
sensors. Specifically, during the simulation, we assumed that there are $%
n_{x}$ satellites located in the platform on each of the semi-axes. Then, it
is obvious that the ensemble of measured coordinates consists of%
\begin{equation}
N_{x}=n_{x}^{6}  \label{67}
\end{equation}%
elements. At the same time, on each of the axes, we have arranged $%
n_{v}=n_{x}^{2}$ velocimeters so that the ensemble of velocity measurements
consists of the same number of elements%
\begin{equation}
N_{v}=n_{v}^{3}=N_{x}.  \label{68}
\end{equation}

\begin{figure*}[t]
\includegraphics[width=1.9\columnwidth]{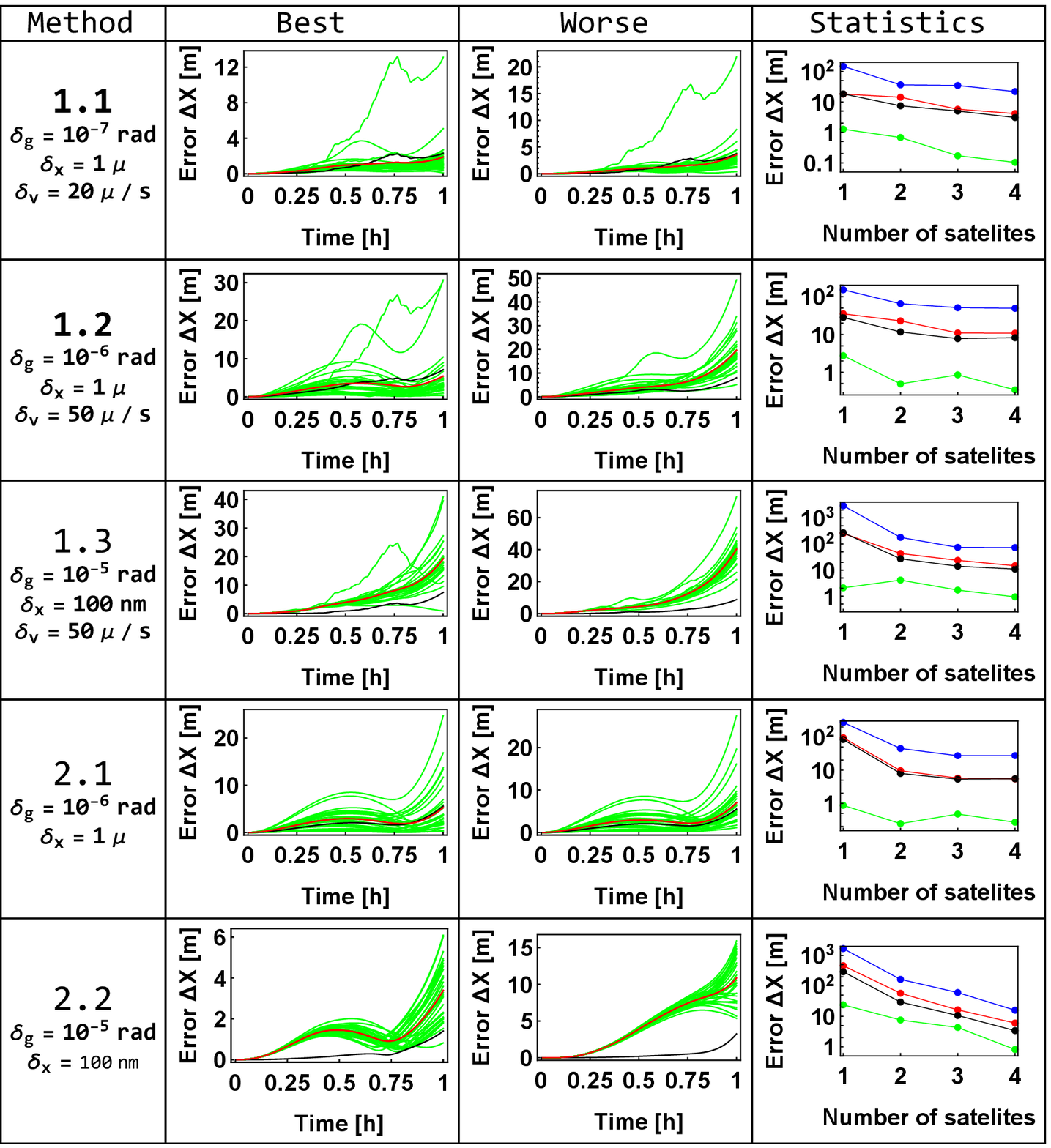}
\caption{Five LPS-N methods. Results of
simulations. For the same 30 trajectories as in FIG. \protect\ref{f1} one
generated a navigation process. This was done for 11 sets of pseudo-random
errors of satellites and velocimeters. As a result, one has ensemble of 330 $%
\Delta X\left( t\right) -$dependences. The average over this ensemble errors
for $t_{n}=1$hour of the navigation are shown by red dots in the 4th column.
In the same plots, the blue, green and black dots show the maximum and
minimum values of the error and SD for the entire ensemble, respectively.
The $\Delta X\left( t\right) -$dependences for 30 platform trajectories with
a given set of sensors' errors are shown in the 2nd and 3rd columns. They
contain plots for those sets of sensor errors at which one of the $\Delta
X\left( t\right) $ curves reaches the minimum (best) or maximum (worst)
value of $\Delta X\left( t_{n}\right) $. \label{f5}  }
\end{figure*}

The simulation showed that the navigation error on the one side is not
sensitive to the position of satellites and velocimeters, and on the other
side depends on the values of particles' launch velocities. We have not been
able to find any qualitative or quantitative criteria to minimize this
dependence. We emphasize that we are talking about a 9-dimensional velocity
space $\left\{ \mathbf{v,v}^{\prime }\mathbf{,v}^{\prime \prime }\right\} $.
Therefore, we randomly selected the velocity components in a given range,
gradually increasing the size of this range until, at $n_{x}=1$, the
navigation accuracy drops to the value of $\Delta X\left( t_{n}\right) \sim
100$m. We expected that with an increase in the number of satellites along
each of the semi-axes $n_{x}$ and a simultaneous increase in the number of
velocimeters $n_{v}=n_{x}^{2}$, the error of the navigation will drop to the
desired level of $\Delta X\left( t_{n}\right) \sim 10$m. The
simulation confirmed this expectation (see the red curves in the fourth
column in Fig. \ref{f5}). The values of launching velocities chosen for the
simulation are given in the table \ref{t1}.

We assume that a navigation cuboid with dimensions $\left\{
L_{N},L_{E},L_{D}\right\} $ is installed in the (NED)-coordinate system of
the platform, which is shielded from all fields and in which a vacuum is
maintained sufficient for one to believe that particles move freely only
under the action of the Earth gravitational field. The minimal sizes of the
cuboid at which there will be no collision of particles with the cuboid
walls are also given in the table \ref{t1}. Since in the 2nd order LPS-N one
observes the particle motion during the time $2T$, then the sizes of the
cuboid are larger than for the 1st order LPS-N at the given $T.$

\begin{table}[tbp]
\caption{Launching velocities chosen for navigation simulations and
corresponding sizes of the navigation cuboids.}
\label{t1}%
\begin{tabular}{|c|c|c|}
\hline
\begin{tabular}{l}
M \\ 
e \\ 
t \\ 
h \\ 
o \\ 
d%
\end{tabular}
& 
\begin{tabular}{c}
Launching \\ 
velocities \\ 
\lbrack m/s]%
\end{tabular}
& 
\begin{tabular}{c}
Minimal \\ 
cuboid \\ 
size \\ 
\lbrack m]%
\end{tabular}
\\ \hline
1.1 & $%
\begin{array}{c}
v_{N}=-0.40 \\ 
v_{E}=-0.40 \\ 
v_{D}=-0.99%
\end{array}%
$ & $%
\begin{array}{c}
L_{N}=0.12 \\ 
L_{E}=0.12 \\ 
L_{D}=0.20%
\end{array}%
$ \\ \hline
1.2 & $%
\begin{array}{ccc}
v_{N}=-0.40 & v_{N}^{\prime }=-0.40 & v_{N}^{\prime \prime }=0.40 \\ 
v_{E}=-0.40 & v_{E}^{\prime }=0.40 & v_{E}^{\prime \prime }=-0.40 \\ 
v_{D}=-1.55 & v_{D}^{\prime }=-3.04 & v_{D}^{\prime \prime }=-1.79%
\end{array}%
$ & $%
\begin{array}{c}
L_{N}=0.12 \\ 
L_{E}=0.12 \\ 
L_{D}=0.49%
\end{array}%
$ \\ \hline
1.3 & $%
\begin{array}{ccc}
v_{N}=2.06 & v_{N}^{\prime }=-1.96 & v_{N}^{\prime \prime }=0.13 \\ 
v_{E}=-0.29 & v_{E}^{\prime }=0.76 & v_{E}^{\prime \prime }=0.67 \\ 
v_{D}=-4.54 & v_{D}^{\prime }=-2.00 & v_{D}^{\prime \prime }=1.15%
\end{array}%
$ & $%
\begin{array}{c}
L_{N}=1.2 \\ 
L_{E}=0.3 \\ 
L_{D}=1.7%
\end{array}%
$ \\ \hline
2.1 & $%
\begin{array}{ccc}
v_{N}=1.77 & v_{N}=1.77 & v_{N}=-1.77 \\ 
v_{E}=1.77 & v_{E}^{\prime }=-1.77 & v_{E}^{\prime }=1.77 \\ 
v_{D}=2.37 & v_{D}^{\prime }=2.05 & v_{D}^{\prime \prime }=1.51%
\end{array}%
$ & $%
\begin{array}{c}
L_{N}=2.1 \\ 
L_{E}=2.1 \\ 
L_{D}=2.4%
\end{array}%
$ \\ \hline
2.2 & $%
\begin{array}{ccc}
v_{N}=2.94 & v_{N}^{\prime }=2.94 & v_{N}^{\prime \prime }=-2.94 \\ 
v_{E}=2.94 & v_{E}^{\prime }=-2.94 & v_{E}^{\prime \prime }=2.94 \\ 
v_{D}=0.99 & v_{D}^{\prime }=-5.61 & v_{D}^{\prime \prime }=-2.97%
\end{array}%
$ & $%
\begin{array}{c}
L_{N}=3.6 \\ 
L_{E}=3.6 \\ 
L_{D}=3.95%
\end{array}%
$ \\ \hline
\end{tabular}%
\end{table}

To illustrate the LPS-N, the build up in time of the navigation errors when
the platform moves along the trajectory 20 (see Fig. \ref{f4}) is shown in
Fig. \ref{f6}. Although the various plots on this and next figures may be
difficult to read at standard magnification, they can be read easily using
the zoom feature when read online in PDF format. The navigation method 1.3
is used here. In this method, the largest number of platform parameters are
restored, the orientation of the gimbal, then sequentially the rotation rate
of the gimbal, the velocity and, finally, the position of the platform. At
the same time, only an error in restoring orientation is caused with the
accuracy of measuring the difference in the coordinates of particles
launched at different velocities. Errors in restoring the subsequent
parameter are related both to the quality of satellites and velocimeters,
and to the accuracy of restoring previous parameters. It is for this reason
that the error in Fig. \ref{f6}$b^{\prime }$ is 5 orders of magnitude larger
than the error in Fig. \ref{f2}.

\begin{figure}[!t]
\includegraphics[width=8cm]{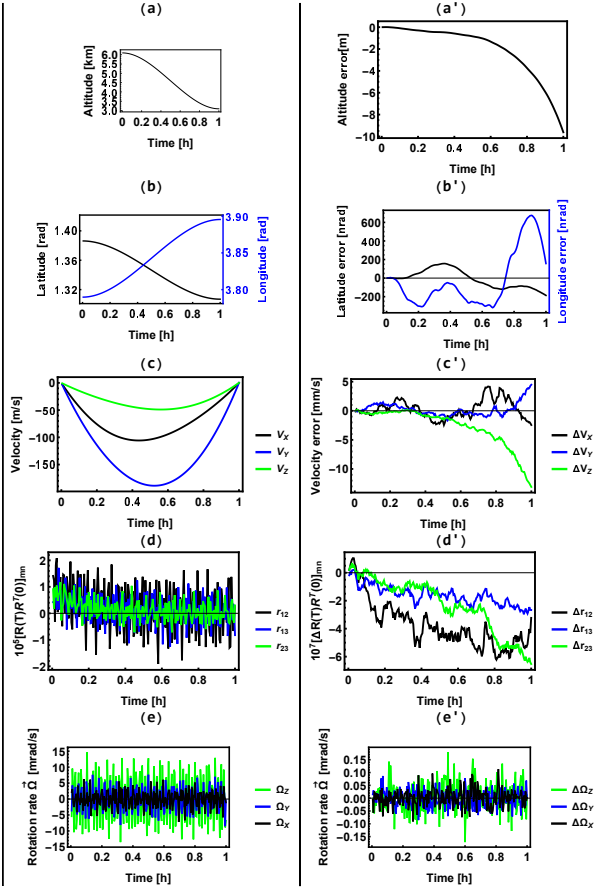}
\caption{The time dependence of the parameters of the platform motion and rotation
(plots $i$) and the errors of their restoring.(plots $i^{\prime }$), where $%
i=a,...,e$. Trajectory 20 (see fig. \protect\ref{f4}). LPS-N method 1.3.
Plots $\left\{ a,a^{\prime }\right\} $ for the altitude of the platform
origine trajectory, plots $\left\{ b,b^{\prime }\right\} $ for the latitude
and longitude of this trajectory, $\left\{ c,c^{\prime }\right\} $ for the
platform velocity $\mathbf{V}\left( t\right) $, $\left\{ d,d^{\prime
}\right\} $ for the matrix elements of the rotation matrix $r\left( t\right)
=R\left( t\right) R^{T}\left( 0\right) $, $\left\{ e,e^{\prime }\right\} $
for the platform rotation rate $\mathbf{\Omega }\left( t\right) .$}
\label{f6}
\end{figure}

The simulation results are summarized in the table in Fig. \ref{f5}.

The five rows of this table correspond to three 1st order LPS-N methods
(1.1, 1.2, and 1.3) and two 2nd order LPS-N methods (2.1 and 2.2). One sees
that in method 1.1, when it is assumed that the quality of the gimbal is
sufficient to neglect the rotation at all, in order to achieve the ultimate
goal of navigation (\ref{6}), it is necessary to use a gimbal with a rms of $%
\delta _{g}=10^{-7}rad$ and velocimeters with an accuracy of $\delta
_{v}=20\mu /s$. The requirements can be mitigated for the gimbal by an order
of magnitude, and for the velocimeters by a factor of 2.5 if one uses method
1.2, when only the angular rate of the gimbal rotation is restored.
Moreover, one can weaken the requirement for a gimbal by an order of
magnitude in the method 1.3, when both the rate of rotation and the
orientation of the gimbal are restored, but satellites with an rms error of $%
\delta _{x}=100$nm are needed. We have verified that with another random
choice of trajectories and orientations, the navigation error $\Delta
X\left( t\right) $ remains the same in the order of its magnitude.

If the 2nd order difference (\ref{3b}) is used, then, due to its low
sensitivity to the platform velocity, one may not use the velocimeters at
all. For both 2nd order LPS-N methods, 2.1 and 2.2, we verified that even
the use of ideal velocimeters, with zero error of velocity measurement, has
no effect on the navigation error $\Delta X\left( t\right) $. In addition,
the lever-arm term disappears, when one launches all three particles from
the same point [see Eq. (\ref{59})], and therefore the platform rotation
rate may not be restored. Finally, with the 2nd order LPS-N , the
requirements for the quality of gimbal are weakened by an order of magnitude
(compare the 4th and 1st rows in the table in Fig. \ref{f5}).

In conclusion, the simulation showed that the LPS can provide navigation
accuracy in 1 hour of the order of 10m, if a gimbal with rms $\delta
_{g}=10^{-6}$rad and satellites, distance sensors, with an error $\delta
_{x}=1\mu $ are used.

\textbf{However, the reverse formulation of the question is also possible.
If the accuracy of the satellites and velocimeters, the rms of gimbal, and
the navigation box sizes are given, then our code will allow us to determine
the expected accuracy of LPS-navigation }$\Delta X$\textbf{\ for a given
time }$t_{n}.$

\acknowledgments The author is grateful to Drs. K. Tintschurin, B. Young, I.
Teper, and A. Zorn for the fruitful discussions.

\appendix

\section{\label{A1}Particle's position and velocity measurements in the
platform frame}

\subsection{\label{A1.1}LPS - measurement of the position}

Let's assume that there are 3 satellites installed on the platform at the
points $\mathbf{s}_{i}=\left( x_{i},y_{i},z_{i}\right) ^{T}$, which measure
the distances to a given point $\mathbf{x}=\left( x,y,z\right) ^{T},$ $%
r_{i}=\left\vert \mathbf{x}-\mathbf{s}_{i}\right\vert $. Then for the point
coordinates one has \cite{c17-1} 
\begin{subequations}
\label{a1}
\begin{gather}
z=\left( -b\pm \sqrt{b^{2}-ac}\right) /a,  \label{a1a} \\
x=b_{0x}+b_{1x}z,  \label{a1b} \\
y=b_{0y}+b_{1y}z,  \label{a1c} \\
b=b_{0x}b_{1x}+b_{0y}b_{1y}-b_{1x}x_{1}-b_{1y}y_{1}-z_{1},  \label{a1d} \\
a=1+b_{1x}^{2}+b_{1y}^{2},  \label{a1e} \\
c=\rho _{1}^{2}-r_{1}^{2}+b_{0x}^{2}+b_{0y}^{2}-2\left(
b_{0x}x_{1}+b_{0y}y_{1}\right) ,  \label{a1f} \\
b_{0x}=\left( a_{2}y_{31}-a_{3}y_{21}\right) /\Delta ,  \label{a1g} \\
b_{1x}=\left( z_{31}y_{21}-z_{21}y_{31}\right) /\Delta ,  \label{a1h} \\
b_{0y}=\left( x_{21}a_{3}-x_{31}a_{2}\right) /\Delta ,  \label{a1i} \\
b_{1y}=\left( x_{31}z_{21}-x_{21}z_{31}\right) /\Delta ,  \label{a1j} \\
\Delta =x_{21}y_{31}-y_{21}x_{31},  \label{a1k} \\
\rho _{i}=\left\vert \mathbf{s}_{i}\right\vert ,  \label{a1l} \\
a_{i}=\dfrac{1}{2}\left( \rho _{i}^{2}-\rho
_{1}^{2}-r_{i}^{2}+r_{1}^{2}\right) ,  \label{a1m} \\
\left( x_{ik},y_{ik},z_{ik}\right) =\left( \mathbf{s}_{i}-\mathbf{s}%
_{k}\right) ^{T}.  \label{a1n}
\end{gather}

\subsection{\label{A1.2}Velocity measurement}

For the 1st order LPS-N, one must measure the velocity $\mathbf{v}$ of a
particle in the platform frame at a given point $\mathbf{x}$. If 3
velocimeters, located at points $\mathbf{s}_{vi}$, measure the projection of
the particle velocity 
\end{subequations}
\begin{equation}
v^{\left( i\right) }=\mathbf{n}_{i}\cdot \mathbf{v},  \label{b1}
\end{equation}%
where

\begin{subequations}
\begin{equation}
\mathbf{n}_{i}=\dfrac{\mathbf{x}-\mathbf{s}_{vi}}{\left\vert \mathbf{x}-%
\mathbf{s}_{vi}\right\vert }.  \label{b2}
\end{equation}%
then for the particle velocity one gets 
\end{subequations}
\begin{equation}
\mathbf{v}=\underline{n}^{-1}\left( 
\begin{array}{c}
v^{\left( 1\right) } \\ 
v^{\left( 2\right) } \\ 
v^{\left( 3\right) }%
\end{array}%
\right) ,  \label{b3}
\end{equation}%
where the matrix $\underline{n}$ is given by%
\begin{equation}
\underline{n}=\left( 
\begin{array}{c}
\mathbf{n}_{1}^{T} \\ 
\mathbf{n}_{2}^{T} \\ 
\mathbf{n}_{3}^{T}%
\end{array}%
\right) .  \label{b4}
\end{equation}

\section{\label{A2}Recurrent relations}

To decompose%
\begin{equation}
\sin \delta =\dsum_{s=0}^{\infty }\dfrac{\left( -1\right) ^{s}\sigma ^{2s+1}%
}{\left( 2s+1\right) !}\left( \dfrac{\delta }{\sigma }\right) ^{2s+1}
\label{c1}
\end{equation}%
one needs to get a series for a degree%
\begin{equation}
\left( \dfrac{\delta }{\sigma }\right) ^{p}=p!\sum_{v=0}^{\infty }\delta
_{v}^{\left( p\right) }\sigma ^{v}  \label{c2}
\end{equation}%
To calculate the coefficient $\delta _{v}^{\left( p\right) },$ one truncates
the row (\ref{55}) with the first $\left( \nu +1\right) $ terms%
\begin{equation}
\delta =\sigma \sum_{i=1}^{\nu +1}\delta ^{\left( i\right) }\sigma ^{i-1}.
\label{c3}
\end{equation}%
Then%
\begin{eqnarray}
\left( \delta /\sigma \right) ^{p} &=&p!\dsum_{r_{1}=0}^{p}\ldots
\dsum_{r_{\nu +1}=0}^{p}\delta _{p,\dsum_{i=1}^{\nu +1}r_{i}}  \notag \\
&&\times \sigma ^{\dsum_{i=1}^{\nu +1}r_{i}\left( i-1\right)
}\dprod\limits_{i=1}^{\nu +1}\dfrac{\left[ \delta ^{\left( i\right) }\right]
^{r_{i}}}{r_{i}!},  \label{c4}
\end{eqnarray}%
where $\delta _{ik}$ is the Kronecker symbol, and so

\begin{eqnarray}
\delta _{v}^{\left( p\right) } &=&\dsum_{r_{1}=0}^{p}\ldots \dsum_{r_{\nu
+1}=0}^{p}\delta _{p,\dsum_{i=1}^{\nu +1}r_{i}}  \notag \\
&&\times \delta _{\nu ,\dsum_{i=1}^{\nu +1}r_{i}\left( i-1\right)
}\dprod\limits_{i=1}^{\nu +1}\dfrac{\left[ \delta ^{\left( i\right) }\right]
^{r_{i}}}{r_{i}!}.  \label{c5}
\end{eqnarray}

Substituting the decomposition (\ref{c2}) in Eq. (\ref{c1}), one finds that 
\begin{subequations}
\label{c6}
\begin{eqnarray}
\sin \delta &\equiv &\dsum_{s=0}^{\infty }u_{s}\sigma ^{s+1},  \label{c6a} \\
u_{s} &=&\dsum_{s^{\prime }=0}^{\left[ s/2\right] }\left( -1\right)
^{s^{\prime }}\delta _{s-2s^{\prime }}^{\left( 2s^{\prime }+1\right) },
\label{c6b}
\end{eqnarray}%
where $\left[ a\right] $ is the Entire part of a. Similarly, one gets

\end{subequations}
\begin{subequations}
\label{c7}
\begin{gather}
1-\cos \delta =\dsum_{s=0}^{\infty }c_{s}\sigma ^{s+2},  \label{c7a} \\
c_{s}=\left\{ 
\begin{array}{c}
\dsum_{s^{\prime }=0}^{\left[ s/2\right] }\left( -1\right) ^{s^{\prime
}}\delta _{s-2s^{\prime }}^{\left( 2s^{\prime }+2\right) },\text{ for }s\geq
0 \\ 
0,\text{ for }s<0%
\end{array}%
\right. .  \label{c7b}
\end{gather}%
Multiplying the power series in Eq. (\ref{53}) one comes to the following
equation for the coefficients

\end{subequations}
\begin{equation}
\dsum_{s_{1}=0}^{s-1}\mathbf{w}_{is_{1}}-\mathbf{a}_{i}c_{s-2}+%
\dsum_{s_{1}=0}^{s-2}c_{s_{1}}\dsum_{s_{2}=0}^{s-s_{1}-2}\mathbf{n}^{\left(
s_{2}\right) }\left( \mathbf{n}^{\left( s-s_{1}-2-s_{2}\right) }\mathbf{a}%
_{i}\right) =\dfrac{\mathbf{b}_{i}}{\sigma }\delta _{s1}.  \label{ccc8}
\end{equation}%
where%
\begin{equation}
\mathbf{w}_{is_{1}}=u_{s_{1}}\mathbf{a}_{i}\times \mathbf{n}^{\left(
s-s_{1}-1\right) }  \label{c8.1}
\end{equation}%
It follows from Eqs. (\ref{c6b}, \ref{c5}, \ref{56a}) that%
\begin{equation}
u_{0}=1  \label{c9}
\end{equation}%
and for $s=1$ one gets the equation%
\begin{equation}
\mathbf{a}_{i}\times \mathbf{n}^{\left( 0\right) }=\dfrac{\mathbf{b}_{i}}{%
\sigma }  \label{c10}
\end{equation}%
the solution of which, taking into account the normalization condition, $%
\left\vert \mathbf{n}^{\left( 0\right) }\right\vert =1$, coincides with (\ref%
{56}).

The coefficients of the our interest, $\left\{ \mathbf{n}^{\left( s-1\right)
},\delta ^{\left( s\right) }\right\} $ are contained only in the terms $%
\mathbf{w}_{i0}$ and $\mathbf{w}_{i\left( s-1\right) }$. For $\mathbf{w}%
_{i0} $ from Eq. (\ref{c9}) one has,%
\begin{equation}
\mathbf{w}_{i0}=\mathbf{a}_{i}\times \mathbf{n}^{\left( s-1\right) },
\label{c11}
\end{equation}%
and from Eqs. (\ref{c6b}, \ref{c10}) one gets%
\begin{equation}
\mathbf{w}_{i\left( s-1\right) }=\left( \delta _{s-1}^{\left( 1\right)
}+q_{s}\right) \dfrac{\mathbf{b}_{i}}{\sigma }  \label{c12}
\end{equation}%
where%
\begin{equation}
q_{s}\equiv \dsum_{s^{\prime }=1}^{\left[ \left( s-1\right) /2\right]
}\left( -1\right) ^{s^{\prime }}\delta _{s-1-2s^{\prime }}^{\left(
2s^{\prime }+1\right) }.  \label{c13}
\end{equation}%
From Eq. (\ref{c5}) for $p=1$ and $\nu =s-1$, one sees that $\delta
_{s-1}^{\left( 1\right) }$ contains only one term at $\left\{ r_{1}\ldots
r_{s}\right\} =\left\{ 0\ldots 01\right\} $, which is equal to%
\begin{equation}
\delta _{s-1}^{\left( 1\right) }=\delta ^{\left( s\right) }.  \label{c14}
\end{equation}%
Then, with $s>1$, one can rewrite Eq. (\ref{ccc8}) as 
\begin{subequations}
\label{c15}
\begin{gather}
\mathbf{a}_{i}\times \mathbf{n}^{\left( s-1\right) }=-\left( \delta ^{\left(
s\right) }+q_{s}\right) \dfrac{\mathbf{b}_{i}}{\sigma }+\mathbf{b}%
_{i}^{\left( s\right) },  \label{c15a} \\
\mathbf{b}_{i}^{\left( s\right) }=\mathbf{a}_{i}c_{s-2}-\dsum_{s_{1}=1}^{s-2}%
\mathbf{w}_{is_{1}}  \notag \\
-\dsum_{s_{1}=0}^{s-2}c_{s_{1}}\dsum_{s_{2}=0}^{s-s_{1}-2}\mathbf{n}^{\left(
s_{2}\right) }\left( \mathbf{n}^{\left( s-s_{1}-2-s_{2}\right) }\cdot 
\mathbf{a}_{i}\right) .  \label{c15b}
\end{gather}

This is an equation of the same type as Eq. (\ref{41}) for the speed of
rotation and Eq. (\ref{50}) for the angle of rotation $\mathbf{\delta }$. By
analogy with those equations, the solution of Eq. (\ref{15a}) is 
\end{subequations}
\begin{equation}
\mathbf{n}^{\left( s-1\right) }=-\left( \delta ^{\left( s\right)
}+q_{s}\right) \mathbf{n}^{\left( 0\right) }+\mathbf{f}\left( \mathbf{a}_{i},%
\mathbf{b}_{i}^{\left( s\right) }\right) ,  \label{c16}
\end{equation}%
where the function $\mathbf{f}$ is given by Eq. (\ref{45b}). Let us consider
the normalization condition $\mathbf{n}^{2}=1.$ It means that%
\begin{eqnarray}
&&2\left[ n^{\left( 0\right) }\cdot \mathbf{n}^{\left( s-1\right)
}+\dsum_{s_{1}=1}^{\left[ s/2\right] -1}\mathbf{n}^{s_{1}}\mathbf{n}%
^{s-s_{1}-1}\right]   \notag \\
&=&-\delta _{s,2\left[ s/2\right] +1}\left[ \mathbf{n}^{\left( \left(
s-1\right) /2\right) }\right] ^{2}  \label{c17}
\end{eqnarray}%
Substituting here. Eq. (\ref{c16}) one arrives to the expression for the
coefficient $\delta ^{\left( s\right) },$%
\begin{gather}
\delta ^{\left( s\right) }=-q_{s}+\mathbf{n}^{\left( 0\right) }\mathbf{f}%
\left( \mathbf{a}_{i},\mathbf{b}_{i}^{\left( s\right) }\right) +  \notag \\
\dsum_{s_{1}=1}^{\left[ s/2\right] -1}\mathbf{n}^{s_{1}}\mathbf{n}%
^{s-s_{1}-1}+\dfrac{1}{2}\delta _{s,2\left[ s/2\right] +1}\left[ \mathbf{n}%
^{\left( \left( s-1\right) /2\right) }\right] ^{2}.  \label{c18}
\end{gather}

The expressions (\ref{c16}, \ref{c18}) are the desired recurrent relations.

Consider them for the lowest orders. If $s=2,$ then from the Eqs..(\ref{c7a}%
, \ref{c13}) it is obvious that $c_{0}=\dfrac{1}{2}$ and $q_{2}=0,$ which
means that%
\begin{equation}
\mathbf{b}_{i}^{\left( 2\right) }=\dfrac{1}{2}\left[ \mathbf{a}_{i}-\mathbf{n%
}^{\left( 0\right) }\left( \mathbf{n}^{\left( 0\right) }\cdot \mathbf{a}%
_{i}\right) \right] .  \label{c19}
\end{equation}%
and, therefore, 
\begin{subequations}
\label{c20}
\begin{eqnarray}
\delta ^{\left( 2\right) } &=&\mathbf{n}^{\left( 0\right) }\cdot \mathbf{f}%
\left( \mathbf{a}_{i},\mathbf{b}_{i}^{\left( 2\right) }\right) ,
\label{c20a} \\
\mathbf{n}^{\left( 1\right) } &=&-\delta ^{\left( 2\right) }\mathbf{n}%
^{\left( 0\right) }+\mathbf{f}\left( \mathbf{a}_{i},\mathbf{b}_{i}^{\left(
2\right) }\right) .  \label{c20b}
\end{eqnarray}

For $s=3$, one consequently finds that 
\end{subequations}
\begin{subequations}
\label{c21}
\begin{eqnarray}
c_{1} &=&u_{1}=\delta ^{\left( 2\right) },  \label{c21a} \\
\mathbf{w}_{i1} &=&\delta ^{\left( 2\right) }\mathbf{a}_{i}\times \mathbf{n}%
^{\left( 1\right) },  \label{c21b}
\end{eqnarray}%
after that 
\end{subequations}
\begin{eqnarray}
\mathbf{b}_{i}^{\left( 3\right) } &=&\delta ^{\left( 2\right) }\left[ 
\mathbf{a}_{i}-\mathbf{a}_{i}\times \mathbf{n}^{\left( 1\right) }-\mathbf{n}%
^{\left( 0\right) }\left( \mathbf{n}^{\left( 0\right) }\mathbf{a}_{i}\right) %
\right]  \notag \\
&&-\dfrac{1}{2}\left( \mathbf{n}^{\left( 0\right) }\left( \mathbf{n}^{\left(
1\right) }\cdot \mathbf{a}_{i}\right) +\mathbf{n}^{\left( 1\right) }\left( 
\mathbf{n}^{\left( 0\right) }\cdot \mathbf{a}_{i}\right) \right) .
\label{c22}
\end{eqnarray}%
And since $q_{3}=-\dfrac{1}{6}$, then%
\begin{equation}
\delta ^{\left( 3\right) }=\dfrac{1}{6}+\mathbf{n}^{\left( 0\right) }\cdot 
\mathbf{f}\left( \mathbf{a}_{i},\mathbf{b}_{i}^{\left( 3\right) }\right) +%
\dfrac{1}{2}\mathbf{n}^{\left( 1\right) 2},  \label{c23}
\end{equation}%
and, finally,%
\begin{equation}
\mathbf{n}^{\left( 2\right) }=\mathbf{f}\left( \mathbf{a}_{i},\mathbf{b}%
_{i}^{\left( 3\right) }\right) -\mathbf{n}^{\left( 0\right) }\left[ \mathbf{n%
}^{\left( 0\right) }\cdot \mathbf{f}\left( \mathbf{a}_{i},\mathbf{b}%
_{i}^{\left( 3\right) }\right) +\dfrac{1}{2}\mathbf{n}^{\left( 1\right) 2}%
\right] .  \label{c24}
\end{equation}

\section{\label{A3}Earth gravitational field}

We consider navigation in the gravitational field of a rotating Earth. This
field consists of normal and anomaly parts%
\begin{equation}
U\left( \mathbf{X}\right) =U_{n}\left( \mathbf{X}\right) +U_{a}\left( 
\mathbf{X}\right)   \label{cc1}
\end{equation}%
For the potential of the normal component, one has \cite{c17} 
\begin{subequations}
\label{cc2}
\begin{align}
U_{n}\left( \mathbf{X}\right) & =-\dfrac{GM}{X}\left( 1-\sum_{m=1}^{\infty
}J_{2m}\left( \dfrac{a_{E}}{X}\right) ^{2m}P_{2m}\left( \cos \theta \right)
\right) ,  \label{cc2a} \\
J_{2m}& =\left( -1\right) ^{m+1}\dfrac{3e^{2m}}{\left( 2m+1\right) \left(
2m+3\right) }\left( 1-m\left( 1-\dfrac{5J_{2}}{e^{2}}\right) \right) ,
\label{cc2b}
\end{align}%
where $GM=3.986004418\cdot 10^{14}$m$^{\text{3}}$s$^{\text{-2}}$ is the
geocentric gravitational constant, $a=6.37813659\cdot 10^{6}$m is the
semimajor axis of the geoid, $e=\sqrt{f\left( 2-f\right) }$ is the first
eccentricity, $f$ is polar flattening $\left( 1/f=298.25765\right) $, $%
J_{2}=1.0826267\cdot 10^{-3}$ is a dynamic form factor of the Earth \cite%
{c18}, $P_{n}\left( x\right) $ is the Legendre polynomial, $\theta $ is the
polar angle of the vector $\mathbf{X}$. From here one derives the following
expressions for the normal components of the gravitational acceleration and
the gravity-gradient tensor of the Earth 
\end{subequations}
\begin{subequations}
\label{cc3}
\begin{eqnarray}
\mathbf{g}_{nE}\left( \mathbf{X}\right)  &=&\mathbf{g}=-\dfrac{GM}{x^{2}}%
\left( b\mathbf{n}+b_{1}\mathbf{z}\right) ;,  \label{cc3a} \\
\underline{\Gamma }_{n}\left( \mathbf{X}\right)  &=&\dfrac{GM}{x^{3}}\left(
-b\delta _{ik}+b_{2}n_{i}n_{k}+b_{3}\left( n_{i}z_{k}+z_{i}n_{k}\right)
\right.   \notag \\
&&\left. +b_{4}z_{i}z_{k}\right) ,  \label{cc3b}
\end{eqnarray}%
where 
\end{subequations}
\begin{subequations}
\label{cc3.1}
\begin{gather}
b=1-\dfrac{1}{\sin ^{2}\theta }\dsum_{m=1}^{\infty }J_{2m}\left( \dfrac{a}{x}%
\right) ^{2m}\left[ P_{2m}\left( \cos \theta \right) \left( 2m+1\right.
\right.   \notag \\
\left. \left. -\left( 4m+1\right) \cos ^{2}\theta \right) +2m\cos \theta
P_{2m-1}\left( \cos \theta \right) \right] ,  \label{cc3.1a} \\
b_{1}=\dfrac{1}{\sin ^{2}\theta }\dsum_{m=1}^{\infty }J_{2m}\left( \dfrac{a}{%
x}\right) ^{2m}2m\left[ P_{2m-1}\left( \cos \theta \right) \right.   \notag
\\
\left. -\cos \theta P_{2m}\left( \cos \theta \right) \right] ,
\label{cc3.1b} \\
b_{2}=3-\dfrac{1}{\sin ^{4}\theta }\dsum_{m=1}^{\infty }J_{2m}\left( \dfrac{a%
}{x}\right) ^{2m}\left\{ P_{2m}\left( \cos \theta \right) \left[ \left(
3+2m\right) \right. \right.   \notag \\
\left. \times \left( 2m+1\right) -\cos ^{2}\theta \left(
20m^{2}+28m+6\right) \right]   \notag \\
\left. +\cos ^{4}\theta \left( 16m^{2}+16m+3\right) \right] +2m\cos \theta  
\notag \\
\left. P_{2m-1}\left( \cos \theta \right) \left[ 5+4m-\left( 3+4m\right)
\cos ^{2}\theta \right] \right\} ,  \label{cc3.1c} \\
b_{3}=\dfrac{1}{\sin ^{4}\theta }\dsum_{m=1}^{\infty }J_{2m}\left( \dfrac{a}{%
x}\right) ^{2m}2m\left\{ 2P_{2m-1}\left( \cos \theta \right) \left(
m+1\right. \right.   \notag \\
\left. \left. -m\cos ^{2}\theta \right) -\cos \theta P_{2m}\left( \cos
\theta \right) \left[ 4m+3-\left( 4m+1\right) \cos ^{2}\theta \right]
\right\} ,  \label{cc3.1d} \\
b_{4}=\dfrac{1}{\sin ^{4}\theta }\dsum_{m=1}^{\infty }J_{2m}\left( \dfrac{a}{%
x}\right) ^{2m}2m  \notag \\
\times \left\{ \left[ 2m+1-\left( 2m-1\right) \cos ^{2}\theta \right]
P_{2m}\left( \cos \theta \right) \right.   \notag \\
\left. -2\cos \theta P_{2m-1}\left( \cos \theta \right) \right\} 
\label{cc3.1e}
\end{gather}%
$\mathbf{z}$ is a unit vector along the Z-axis.

For the anomaly part of the potential $U_{a}\left( \mathbf{X}\right) ,$ we
could not find an analytical expression. Nevertheless, it is known \cite{c19}
that the rms of anomalous acceleration 
\end{subequations}
\begin{equation}
g_{a}\sim 30\text{mGal}  \label{cc4}
\end{equation}%
and from Plate 5 in Ref. \cite{c20} one can conclude that the rms of the
anomaly part of the gravity-gradient tensor $\Gamma _{a}\sim 100$E. It
follows that the typical size of the anomaly field oscillations in space is%
\begin{equation}
L_{a}\sim g_{a}/\left\vert \Gamma _{a}\right\vert \sim 3\text{km.}
\label{cc5}
\end{equation}%
Unfortunately, the known analytical \cite{c21} or numerical \cite{c21.1,c22}
approaches to the calculation of the gravitational field in the space around
the Earth, including its anomalous part, cannot be used in our calculations,
since they assume a known field on the Earth's surface. Therefore, for our
code, we assumed that the anomalous potential consists of the sum of
sinusoidal terms%
\begin{equation}
U_{a}\left( \mathbf{X}\right) =\sum_{j=1}^{n_{a}}h_{aj}\cos \left( \mathbf{k}%
_{aj}\cdot \mathbf{X}\right) ,  \label{cc6}
\end{equation}%
in which the wave vectors were generated by the formula%
\begin{equation}
k_{aji}=\pm 2\pi /L_{a}(1+2r),  \label{cc7}
\end{equation}%
where $r$ is a pseudo-random number uniformly distributed over the interval $%
\left[ 0,1\right] $, and for amplitudes $h_{aj}$, a generator of normal
pseudo-random numbers was used, after which they were normalized by the
condition%
\begin{equation}
g_{a}^{2}=\dfrac{1}{2}\sum_{j=1}^{n_{a}}h_{aj}^{2}k_{aj}^{2}.  \label{cc8}
\end{equation}%
Of course, Eq. (\ref{cc6}) has nothing common with reality [it follows, for
example, that particles move in a hypothetical medium with a density of $%
\rho =-4\pi \triangle U_{a}\left( \vec{r}\right) /G\not=0,$ which can even
be negative. Despite this, we accept it, since at $n_{a}\gg 1$, the
anomalous field has a rms (\ref{cc4}) and a spatial period (\ref{cc5}). In
addition, the representation of the anomalous potential in the form of sums
of harmonic terms can qualitatively correspond to the topographic
undulations observed by the GRACE and GOCE satellites.


\begin{thebibliography}{99}
\bibitem{c1} B Ya Dubetskii, A P. Kazantsev, V P. Chebotayev, V P. Yakovlev,
Interference of atoms in separated optical fields, \textit{Pis'ma Zh. Eksp.
Teor. Fiz.} \textbf{39}, 531 (1984) [\textit{JETP Lett}. \textbf{39}, 649
(1984)].

\bibitem{c1.1} G M Tino and M A Kasevich(ed) 2014 Atom interferometry Proc.
of the Int. School of Physics `Enrico Fermi' vol 188 (IOS Press)

\bibitem{c1.2} B. Canuel et al, Technologies for the ELGAR large scale atom
interferometer array, arXiv:\href{https://arxiv.org/abs/2007.04014}{%
2007.04014v1 [physics.atom-ph]}

\bibitem{c1.3} Ming-Sheng Zhan et al, ZAIGA: Zhaoshan long-baseline atom
interferometer gravitation antenna, Int. J. Mod. Phys. D \href{https://doi.org/10.1142/S0218271819400054}%
{\textbf{29}, 194005 (2020).}

\bibitem{c1.4} L. Badurina et al, AION: an atom interferometer observatory
and network, J. Cosmol. Astropart. Phys. \href{https://doi.org/10.1088/1475-7516/2020/05/011}%
{JCAP05(2020)011}

\bibitem{c1.5} B. Battelier et al, Exploring the Foundations of the Universe
with Space Tests of the Equivalence Principle, arXiv:\href{https://arxiv.org/abs/1908.11785}%
{1908.11785v3 [physics.space-ph]}

\bibitem{c1.5.1} Y. A. El-Neaj et al, AEDGE: Atomic Experiment for Dark
Matter and Gravity Exploration in Space,EPJ Quantum Technology \textbf{\ } 
\href{https://doi.org/10.1140/epjqt/s40507-020-0080-0}{\textbf{7}, 6 (2020).}

\bibitem{c1.5.2} G M. Tino et al, SAGE: A proposal for a space atomic
gravity explorer, Eur. Phys. J. D.\textbf{\ }\href{https://doi.org/10.1140/epjd/e2019-100324-6}%
{\textbf{73}, 228 (2019).}

\bibitem{c1.5.3} M. Abe et al, Matter-wave Atomic Gradiometer
Interferometric Sensor (MAGIS-100), \href{https://arxiv.org/abs/2104.02835}{%
arXiv:2104.02835 [physics.atom-ph]}.

\bibitem{c2} M. Kasevich, S. Chu, Atomic interferometry using stimulated
Raman transitions, Phys. Rev. Lett. \href{https://doi.org/10.1103/PhysRevLett.67.181}%
{\textbf{67}, 181 (1991).}.

\bibitem{c3} P. Asenbaum, C. Overstreet, M. Kim, J. Curti, and M. A.
Kasevich, Atom-Interferometric Test of the Equivalence Principle at the $%
10^{-12}$ Level, Phys. Rev. Lett. \href{https://doi.org/10.1103/PhysRevLett.125.191101}%
{\textbf{125}, 191101 (2020).}

\bibitem{c4} B. Canuel, F. Leduc, D. Holleville, A. Gauguet, J. Fils, A.
Virdis, A. Clairon, N. Dimarcq, Ch. J. Borde, A. Landragin, 6-axis inertial
sensor using cold-atom interferometry, Phys. Rev. Lett. \href{https://doi.org/10.1103/PhysRevLett.97.010402}%
{\textbf{97}, 010402 (2006).}

\bibitem{c5} M. S. Ahmed, D. V. Cuk, Comparison of Different Computation
Methods for Strapdown Inertial Navigation Systems, Sci. Tech. Rev. \textbf{LV%
}, 22 (2005).

\bibitem{c6} C. Jekeli, "Inertial navigation systems with geodetic
applications," de Gruyter, Berlin, New York, 2001.

\bibitem{c7} Aaron Canciani. Integration of cold atom interferometry ins
with other sensors. Master's thesis, Second Lieutenant, Air Force Institute
of Technology (USAF), 2012.

\bibitem{c8} J. Lautier, L. Volodimer, T. Hardin, S. Merlet, M. Lours, F.
Pereira Dos Santos, and A. Landragin, Hybridizing matter-wave and classical
accelerometers, Appl. Phys. Lett. \href{https://doi.org/10.1063/1.4897358}{%
\textbf{105}, 144102 (2014).}

\bibitem{c9} P. Cheiney, L. Fouch\'{e}, S. Templier, F. Napolitano, B.
Battelier, P. Bouyer, and B. Barrett. Navigation-compatible hybrid quantum
accelerometer using a Kalman filter. Phys. Rev. Applied \href{https://doi.org/10.1103/PhysRevApplied.10.034030}%
{\textbf{10}, 034030 (2018).}

\bibitem{c9.1} Y. Wu, J. Guo, X. Feng, L. Q. Chen, C-H. Yuan, W. Zhang,
Atom-light hybrid quantum gyroscope, Phys. Rev. Applied \href{https://doi.org/10.1103/PhysRevApplied.14.064023}%
{\textbf{14}, 064023 (2020).}

\bibitem{c10} X. Wang, A. Kealy, C. Gilliam, S. Haine, J. Close, B. Moran,
K. Talbot, S. Williams, K. Hardman, C. Freier, P. Wigley, A. White, S.
Szigeti, S. Legge, Enhancing Inertial Navigation Performance via Fusion of
Classical and Quantum Accelerometers, \href{https://arxiv.org/abs/2103.09378v1}%
{arXiv:2103.09378v1 [quant-ph]}.

\bibitem{c11} M. A. Kasevich and B. Dubetsky, Kinematic Sensors Employing
Atom Interferometer Phases, US Patent 7,317,184 (2005).

\bibitem{c12} G.E.P. Box, , M.E. Muller, A note on the generation of random
normal deviates, \href{https://projecteuclid.org/journals/annals-of-mathematical-statistics/volume-29/issue-2/A-Note-on-the-Generation-of-Random-Normal-Deviates/10.1214/aoms/1177706645.full}%
{Annals Math. Stat, 29, 610 (1958).}

\bibitem{c13} M. A. Kasevich and B. Dubetsky, The phase of an atom
interferometer as a direct source for precise navigation, private
communication.

\bibitem{c14} N. F. Ramsey, A new molecular beam resonance method, \href{https://doi.org/10.1103/PhysRev.76.996}%
{Phys. Rev. \textbf{76}, 996 (1949).}

\bibitem{c14.1} B. Dubetsky and M. A. Kasevich, Atom interferometer as a
selective sensor of rotation or gravity, Phys. Rev. A \textbf{74}, 023615
(2006).

\bibitem{c14.2} K. Bongs, R. Launay, and M. A. Kasevich, High-order inertial
phase shifts for time-domain atom interferometers, Appl. Phys. B \textbf{84}%
, 599 (2006).

\bibitem{c14.3} J. M. Hogan, D. M. S. Johnson, M. A. Kasevich, Light-pulse
atom interferometry, arXiv:0806.3261 [physics.atom-ph], appear in the
Proceedings of the International Summer School of Physics "Enrico Fermi" on
Atom Optics and Space Physics (Varenna, July 2007).

\bibitem{c14.4} S. Kleinerta, E. Kajaria, A. Rouraa, W. P. Schleich,
Representation-free description of light-pulse atom interferometry including
non-inertial effects, Phys. Rept. 605, 1 (2015).

\bibitem{c15} G. L. Kotkin and V. G. Serbo, "Collection of problems in
classical mechanics," Oxford, New York, Pergamon Press (1971), problem 6.23.

\bibitem{c16} C. K. Chui, G. Chen, "Kalman filtering with real-time
applications," Springer, Berlin, Heidelberg (1999), Sec. 1.1

\bibitem{c16.1} O. Rodrigues, "Des lois g\'{e}om\'{e}triques qui r\'{e}%
gissent les d\'{e}placements d'un syst\`{e}me solide dans l'espace, et de la
variation des coordonn\'{e}es provenant de ces d\'{e}placements consid\'{e}r%
\'{e}s ind\'{e}pendants des causes qui peuvent les produire", Journal de Math%
\'{e}matiques Pures et Appliqu\'{e}es \textbf{5}, 380 (1840).

\bibitem{c17-1} V. S. Shebshaevich, P. P. Dmitriev, N. V. Ivancevich, A. V.
Kalugin, E. G. Kovalevsky, I. V. Kudryavtsev, V. Yu. Kutikov, Yu. B.
Molchanov, Yu. A. Maksyutenko, Setevye sputnikovye radionavigatsionnye
sistemy, RADIO I SVYAZ', Moscow 1993, p. 220 (in Russian).

\bibitem{c17} W.A. Heiskanen and H. Moritz, "Physical Geodesy," W.H. Freeman
and Co. San Francisco (1967), Sec. 2-9

\bibitem{c18} E. Groten, "Parameters of Common Relevance of Astronomy,
Geodesy, and Geodynamics," pp. 134-140 in H. Moritz, Geodetic reference
system 1980, \href{https://doi.org/10.1007/s001900050278}{Journal of
Geodesy, \textbf{74, }128 (2000).}

\bibitem{c19} Committee on Earth Gravity from Space, in Satellite Gravity
and the Geosphere: Contributions to the Study of the Solid Earth and Its
Fluid Envelopes (National Academies Press, Washington, DC, 1997), p. 13.

\bibitem{c20} Y. M. Wang, GSFCO0 mean sea surface, gravity anomaly, and
vertical gravity gradient from satellite altimeter data, \href{https://doi.org/10.1029/2000JC000470}%
{J. Geophys. Res. \textbf{106}, 31167 (2001).}

\bibitem{c21} J. Yu, C. Jekeli \& M. Zhu, Analytical solutions of the
Dirichlet and Neumann boundary-value problems with an ellipsoidal boundary, 
\href{https://doi.org/10.1007/s00190-002-0271-8}{Journal of Geodesy 76,
653--667 (2003). }

\bibitem{c21.1} C. Jekeli, J. K. Lee, J. H. Kwon, Modeling errors in upward
continuation for INS gravity compensation, \href{https://doi.org/10.1007/s00190-006-0108-y}%
{Journal of Geodesy, 81, 297 (2007).}

\bibitem{c22} Y. M. Wang, Geodetic Boundary Value Problems, In Encyclopedia
of Geodesy, E.W. Grafarend (ed.), Springer International Publishing
Switzerland \href{https://doi.org/10.1007/978-3-319-02370-0_42-1}{(Outside
the USA) 2016.}
\end{thebibliography}
\end{document}